\documentclass[11pt,a4paper]{article}

\usepackage[english]{babel}

\pdfoutput=1 

\RequirePackage{ifpdf}
\usepackage{amsmath}
\usepackage{mathtools}

\usepackage{jheppub}
\usepackage{pstricks}
\usepackage[final]{pdfpages}
\usepackage{ifpdf}
\usepackage{slashed}
\usepackage{subfiles}

\usepackage{subcaption}

\usepackage{hyperref}

\usepackage{color}
\usepackage{graphics}
\usepackage{multirow}
\usepackage{etoolbox} 
\usepackage{fixmath}

\usepackage{notoccite}

\usepackage{amsfonts}
\usepackage{float}

\usepackage[normalem]{ulem}
\usepackage{tabularx,siunitx}

\usepackage{tikz}

\newcommand{\oceann}{{\sc Oceann }}
\newcommand{\abiss}{{\sc Abiss}}
\def\seasyde{{\sc SeaSyde}}
\def\amflow{{\sc AMFlow}}
\newcommand{\kira}{{\sc KIRA}}
\newcommand{\oa}{${\cal O}(\alpha)\,$}
\newcommand{\oaa}{${\cal O}(\alpha^2)\,$}
\newcommand{\oaaa}{${\cal O}(\alpha^3)\,$}
\newcommand{\mw}{$\mu_W$}
\newcommand{\mz}{$\mu_Z$}
\newcommand{\mh}{$\mu_H$}
\newcommand{\msbar}{{ \overline{\mathrm{MS}}}}

\definecolor{myblue}{RGB}{0, 51, 102}

\definecolor{darkgreen}{rgb}{0,0.5,0}
\newcommand{\sd}[1]{{\color{darkgreen} #1}}

\def\M{{\cal M}}
\def\unM{\hat{\cal M}}

\def\I{{\cal I}}

\title{Towards the two-loop electroweak corrections to the  Drell-Yan process: the complete fermionic contributions}

\author[a,b]{Tommaso Armadillo,}
\author[c]{Simone Devoto,}
\author[b]{Michele Dradi,}
\author[b]{Alessandro Vicini}

\affiliation[a]{Centre for Cosmology, Particle Physics and Phenomenology (CP3), Université catholique de Louvain, Chemin du Cyclotron, 2, B-1348 Louvain-la-Neuve, Belgium}
\affiliation[b]{Dipartimento di Fisica ``Aldo Pontremoli'',
  University of Milano and INFN, Sezione di Milano, I-20133 Milano, Italy}
\affiliation[c]{Vakgroep Fysica en Sterrenkunde, Universiteit Gent, 9000 Gent, Belgium}

\emailAdd{tommaso.armadillo@uclouvain.be}
\emailAdd{simone.devoto@ugent.be}
\emailAdd{michele.dradi@mi.infn.it}
\emailAdd{alessandro.vicini@mi.infn.it}

\abstract{
We discuss  the production of a lepton pair in quark-antiquark annihilation in the electroweak Standard Model.
We present the   ultraviolet-renormalised and infrared-subtracted finite contribution of the complete set of second-order virtual corrections with a closed fermionic loop, to the cross section of the process $u\bar u\to \mu^+\mu^-$.
The evaluation of these contributions is based on an automated methodology, whose validity is discussed with explicit tests. 
These corrections are one of the building blocks necessary for the simulation of the inclusive lepton-pair production cross section at next-to-next-to-leading order in the electroweak interaction.
}


\begin{document}
\allowdisplaybreaks[4]
\unitlength1cm
\maketitle
\flushbottom

\section{Introduction}
\setcounter{equation}{0}
\label{sec:intro}
After the proof of renormalisability of quantum gauge field theories~\cite{tHooft:1972tcz,tHooft:1972qbu,Becchi:1974md,Becchi:1975nq} and in particular of the electroweak (EW) Standard Model (SM)~\cite{Ross:1973fp}, the study of EW phenomenology at the quantum level has become possible with the complete solution of all the 1-loop scalar Feynman integrals~\cite{tHooft:1978jhc} and with the first calculation of the complete set of first-order quantum corrections to the cross section of the inclusive production of a muon pair in electron-positron annihilation~\cite{Passarino:1978jh}. The calculation of the full set of second-order EW corrections to this process is still missing today and represents a major challenge in Quantum Field Theory.

The 17M of events collected at the $Z$ resonance by the experiments at the CERN LEP and the SLAC SLD colliders have allowed the determination of the couplings of the $Z$ boson to fermions, of its mass and decay width, with per mille precision or better ($10^{-5}$ in the case of the mass)~\cite{ALEPH:2005ab,ParticleDataGroup:2020ssz}.
Their determination from the data has required the inclusion of several classes of radiative corrections, beyond the first-order ones, such as the description of multiple photon and gluon emissions, the dominant fermionic contributions to the gauge-boson self-energies\sd{,} correcting the tree-level propagators\sd{,} and to the gauge boson-fermion-antifermion vertices~\cite{Bardin:1997xq}. The introduction of the $\Delta r$ parameter~\cite{Sirlin:1980nh} in the discussion of the charged-current weak interaction, has been accompanied by the definition of the $\rho$ parameter~\cite{Marciano:1980pb},
expressing  the strength of the neutral current weak interaction in terms of the charged-current one, and of the $\Delta\kappa$ parameter, relevant in the definition of an effective fermionic weak mixing angle at the $Z$ resonance~\cite{Marciano:1980pb}.
These three quantities can be computed
with the inclusion of universal higher-order corrections, thus improving the accuracy of the cross section predictions.
The full set of second-order EW corrections 
to the above parameters has become available in the last two decades, consolidating the theoretical framework necessary to interpret the LEP/SLD data~\cite{Dubovyk:2019szj}. 

The production of a lepton pair is a benchmark process at both lepton and proton colliders.
The latter case is usually referred to as the Drell-Yan (DY) process, and has been an important reference for the development of higher-order calculations in the SM, including
corrections up to the third order in Quantum Chromodynamics (QCD)~\cite{Duhr:2020seh,Duhr:2020sdp,Duhr:2021vwj}, up to the Next-To-Leading Order (NLO) in the EW interaction~\cite{Dittmaier:2001ay,Baur:2004ig,Zykunov:2006yb,Arbuzov:2005dd,CarloniCalame:2006zq,Baur:2001ze,Zykunov:2005tc,CarloniCalame:2007cd,Arbuzov:2007db,Dittmaier:2009cr} and at Next-to-Next-to-Leading-Order (NNLO) with mixed QCD-EW corrections~\cite{Bonciani:2021zzf,Buccioni:2022kgy,Armadillo:2024nwk}.
The possibility at the CERN LHC to study final states at the TeV scale has drawn the attention to another subset of EW quantum corrections, enhanced by the so called EW Sudakov logarithms: the latter become relevant when their argument $q^2/m_V^2\gg1$, with $q^2$ one of the kinematical invariants of the scattering process and $m_V$ a gauge boson mass~\cite{Ciafaloni:1998xg}. The size of these effects, which stem only from the virtual corrections, can become as large as several tens of percent of the Born approximation, and has motivated their study not only at first order, but also their resummation in leading and next-to-leading logarithmic approximation and the study of their role in conjunction with QCD effects~\cite{Denner:2000jv,Denner:2006jr,Denner:2024yut}.

\begin{table}[th]
    \centering
    \begin{tabular}{|c|c|c|}
    \hline
     bin range (GeV)    & \% error with 140 fb$^{-1}$  & \% error with 3 ab$^{-1}$\\
\hline
\hline
91-92    & 0.03 & 0.006\\
120-400  & 0.1~ & 0.02~~\\
400-600  & 0.6~ & 0.13~~\\
600-900  & 1.4~ & 0.30~~\\
900-1300 & 3.2~ & 0.69~~\\
\hline
    \end{tabular}
    \caption{Statistical errors on the measurement of the differential cross section for the production of a muon pair of given invariant mass in proton-proton collisions; values expected at the LHC, with the luminosities at the end of Run-II and at the end of the High-Luminosity phase (estimates based on the cross sections of~\cite{Bonciani:2021zzf}). }
    \label{tab:staterrorsLHC}
\end{table}
\begin{table}[th]
    \centering
    \begin{tabular}{|c|c|c|c|}
    \hline
     $\sqrt{S}$ (GeV)    & luminosity (ab$^{-1}$)  & $\sigma$ (fb) & \% error  \\
\hline
\hline
 ~91  & 150 & $2.17595\cdot 10^{6}$ & 0.0002\\
240  & 5   & 1870.84               & 0.03\\
365  & 1.5 & 787.74                & 0.09\\
\hline
    \end{tabular}
    \caption{Statistical errors on the measurement of the total cross section for the production of a muon pair of given collider energies $\sqrt{S}$  in electron-positron collisions; values expected at the FCC-ee~\cite{deBlas:2022ofj}. }
    \label{tab:staterrorsFCC}
\end{table}
The precision targets prospected at the LHC in its High-Luminosity operation phase, and the one foreseen at the FCC-ee, define an unprecedented challenge to the theoretical predictions needed for phenomenological studies.
We read from Tables \ref{tab:staterrorsLHC} and \ref{tab:staterrorsFCC} that sub-per mille precision will be achievable in the measurement of the cross sections of the production of a muon pair, spanning a very wide range of final-state  invariant masses,  at both kinds of colliders (proton-proton and electron-positron). 
Using the statistical error as the proxy for the ultimate experimental precision (clearly an ideal limit when all the systematics can be neglected),
we can classify the relevance of the different subsets of higher-order SM radiative corrections needed for a statistically meaningful comparison with the data. The statistics at the $Z$ resonance clearly calls for the evaluation of higher-order corrections, at least at full second order in the EW interaction~\cite{FCC:2025lpp}.
The invariant mass region above the gauge boson resonances, relevant for the searches for signals of physics beyond the SM \cite{Armadillo:2026mvp}, will also be measured very precisely.
The several per cent size of the EW Sudakov logarithms contributions~\cite{Jantzen:2005az}, the alternating signs of the logarithmic expansion and the not necessarily small value of the constant term~\cite{Pagani:2021vyk} suggest that the explicit evaluation of the full set of the exact  NNLO EW corrections to the  production of a fermion pair 
is urgent, in order to reduce the residual uncertainties of the theoretical predictions below the current several percent  level in hadron-hadron collisions~\cite{Chiesa:2024qzd}, and to match the expected ultimate experimental precision.

The  challenges that have to be faced to complete the evaluation of the complete second order EW corrections
are of theoretical and computational nature.
The ultraviolet (UV) renormalisation~\cite{Degrassi:2003rw,Actis:2006rb,Actis:2006rc,Denner:2019vbn,Dittmaier:2021loa} in presence of unstable particles and the subtraction of infrared (IR) divergences~\cite{Armadillo:2025mfx} has to be systematically organised, together with a clear treatment of the Dirac algebra in presence of chiral couplings~\cite{'tHooft:1972fi,Breitenlohner:1975hg,Breitenlohner:1976te,Breitenlohner:1977hr,Kreimer:1989ke,Korner:1991sx,Kreimer:1993bh,Heller:2020owb}.
The evaluation of the Feynman integrals is another demanding problem (see e.g.~\cite{Armadillo:2022ugh,Liu:2022chg} and references therein), together with the handling and bookkeeping of a very large number of Feynman diagrams which contribute to the scattering amplitude~\cite{ABISS}.

There is a limited number of examples of physical processes, whose theoretical predictions include the second-order EW effects. The muon decay amplitude studied at zero momentum transfer has been evaluated with the full set of two-loop virtual EW corrections~\cite{Freitas:2002ja,Awramik:2003ee}.
The second-order fermionic EW corrections to the polarised Moeller scattering asymmetry have been presented in~\cite{Du:2019evk}.
A partial set of the second-order EW fermionic corrections to the 
neutral current (NC) DY amplitude has been discussed in~\cite{Freitas:2025vax}.

In this paper we present the evaluation of the complete  subset  of the second-order fermionic EW virtual corrections to the production of a muon pair in quark-antiquark annihilation, for arbitrary kinematics. This subset is defined by the presence of a closed fermionic loop in the Feynman diagrams or in the relevant UV counterterms.
The successful completion of the calculation requires the solution of the above-mentioned  theoretical and computational problems and their implementation and automation in efficient computer packages.
This calculation represents an important step towards the evaluation of the complete set of second-order virtual EW corrections to the DY process.
We take the opportunity to illustrate the intricate structure of the singularity cancellations which take place when evaluating the complete virtual amplitude, providing a validation of our tools.
We also illustrate the tests that support the validity of the evaluation of the finite part of these corrections, as a benchmark for future independent comparisons. We address the open issues that have to be discussed in the combination of the  corrections to the NC DY cross section at NNLO-EW level, including several subtleties which originate from the two-loop virtual sector.

The paper is organised as follows:
in Section \ref{sec:renormalization} we present the general structure of the UV counterterms at second order EW;
in Section \ref{sec:infra} we discuss the subtraction of the IR divergences;
in Section \ref{sec:gamma5} we describe the basic elements of the prescription we adopted to compute the trace of a product of Dirac matrices and $\gamma_5$ in dimensional regularisation;
in Section \ref{sec:amplitude} 
we describe the specific set of contributions considered in the calculation;
in Section \ref{sec:results}
we discuss the main results of our computation and their validity;
eventually, in Section \ref{sec:conclusions}, we draw our conclusions.

\section{Ultraviolet renormalisation}
\label{sec:renormalization}

The SM UV renormalisation up to second order in the electroweak interaction has been discussed in \cite{Denner:1994xt,Denner:2019vbn,Actis:2006ra,Actis:2006rb,Actis:2006rc, Freitas:2002ja} both for the on-shell and the $\overline{MS}$ renormalisation schemes. The electric charge renormalisation at \oaa has been presented in detail in \cite{Degrassi:2003rw,Dittmaier:2021loa}. 
In order to relate the gauge couplings $g$ and $g'$, the vacuum expectation value $v$ of the Higgs doublet and the scalar quartic coupling $\lambda$, to a set of experimental inputs, we
choose the $W$ and $Z$ gauge boson masses, the Higgs mass and the 
electromagnetic coupling.
We consider here the Complex Mass Scheme (CMS) for the definition of the  gauge boson masses \cite{Denner:2005fg}, while we choose the $\msbar$ scheme for the electromagnetic coupling, dubbing this choice ($\alpha_\msbar$,\,\mw,\,\mz,\,\mh) input scheme. In addition, the Yukawa couplings are related to the corresponding fermion masses. The tree-level relations among the bare Lagrangian couplings and bare input parameters read
\begin{align}
    &
    e_0=\frac{g_0 g'_0}{\sqrt{g_0^2+g_0^{'2}}},
    & &
    \mu_{W,0}^2 = \frac{g_0^2 v_0^2}{4},
&&
\mu_{Z,0}^2 = \frac{\left(g_0^2+g_0^{'2}\right) v_0^2}{4},
    \nonumber&\\
    &
    \mu_{H,0}^2 = 2\lambda_0 v_0^2,
    & &
    m_{f,0}=\frac{y_{f,0} v_0}{\sqrt{2}}\,.
    \label{eq:bareinputcouplings}
\end{align}
The expression of the bare Higgs mass in terms of the quartic coupling will be further discussed in Section \ref{sec:Tadpole convention and Goldstone counterterms}. We treat the quantisation of the EW SM in the Background Field Gauge (BFG) approach \cite{Denner:1994xt,Denner:2019vbn}, where the scalar and vector fields of the classical Lagrangian are split into the sum of classical background and quantum fields $\hat{V}+V$, only the latter representing the integration variables of the SM effective action functional integral. The gauge-fixing term of the Lagrangian is hence introduced differently for the background and the quantum  fields;
for the latter it is chosen in such a way to  preserve QED-like Ward Identities for the Green's functions involving a background photon field, even in the complete EW SM. The renormalisation of the theory is achieved with the introduction of a suitable set of counterterms for the bare input parameters
\begin{align} 
    \nonumber
    &
    e_0 = e\ (1 +\delta Z_e )\,,
    & &
    \mu_{W,0}^2=\mu_W^2 + \delta \mu_W^2 \,,
    &&
   \mu_{Z,0}^2=\mu_Z^2 + \delta \mu_Z^2 \,,
    \\
    \label{eq:bareparamct}
    &
    \mu_{H,0}^2=\mu_H^2 + \delta \mu_H^2 \,,
    &&
    m_{f,0} = m_f + \delta m_f \,,
\end{align}
and for the wave function (WF) renormalisation constants of the fermions and all the background fields:
\begin{align}
    \psi^{f}_{L/R,0} &= \left(Z^f_{L/R}\right)^{{1}/{2}} \psi^{f}_{L/R} = \left(1+\delta Z^f_{L/R}\right)^{1/2}\psi^{f}_{L/R} \ , \nonumber \\
     \bar{\psi}^{f}_{L/R,0} &= \left(Z^{\bar{f}}_{L/R}\right)^{{1}/{2}} \bar{\psi}^{f}_{L/R} = \left(1+\delta Z^{\bar{f}}_{L/R}\right)^{1/2} \bar{\psi}^{f}_{L/R} \ , \nonumber \\
     \hat\pi_0&=Z_\pi^{1/2}\,\hat\pi= \left(1+\delta Z_\pi\right)^{1/2}\,\hat\pi \ ,
\end{align}
with $\hat\pi = (\hat{W}^\pm,\hat{H},\hat\chi,\hat{\phi}^\pm)$.
For the neutral gauge bosons, the possibility of a photon-$Z$ mixing leads to the introduction of two additional constants, relating bare, $\hat{A}_0$ and $\hat{Z}_0$  and renormalised, $\hat{A}$ and $\hat{Z}$, background photon and $Z$ fields:
\begin{align}
    &
    \begin{pmatrix} \hat{Z}_0 \\ \hat{A}_0 \end{pmatrix}  = \begin{pmatrix} Z_{ZZ}^{\frac{1}{2}} & \left(Z_{ZA}^{\frac{1}{2}} -1\right) \\
    \left(Z_{AZ}^{\frac{1}{2}}-1\right) & Z_{AA}^{\frac{1}{2}} \end{pmatrix} 
    \begin{pmatrix} \hat{Z} \\ \hat{A} \end{pmatrix}\,.
\label{eq:barefieldct}
\end{align}
The renormalisation of the external fermion fields  is required to ensure the probabilistic interpretation of the squared amplitude.
The renormalisation constants of the background Goldstone bosons and all the quantum fields, including the Faddeev-Popov ghost fields, exactly cancel and do not contribute  to the S matrix. 
The WF renormalisation constants of the background Higgs and gauge bosons fields enter in the expression of the mass counterterms at second order in perturbation theory and thus have to be specified, as discussed in the following Sections.
Since we do not renormalise the gauge parameter, we have in general UV divergent Green's functions. We consider the cancellation of the UV poles at the level of the complete amplitude as a strong check of our implementation of the SM Feynman rules and of the gauge invariance of the amplitude. The rules have been collected in a dedicated model file, starting from the BFG template of {\sc FeynArts} \cite{Hahn:2000kx} and adding the renormalisation counterterms up to \oaa.
After imposing suitable renormalisation conditions, the counterterms can be computed as a perturbative expansion in the electromagnetic coupling constant\footnote{In the full SM we have a double perturbative expansion in the electromagnetic and strong coupling constants. For the sake of simplicity, since we discuss second-order EW corrections, we focus only on the electromagnetic part.}, starting at \oa.

\subsection{Vertex functions and renormalisation conditions}

Using the transverse and longitudinal projectors $T^{\mu \nu} \equiv g^{\mu \nu} - k^{\mu}k^{\nu}/k^2$ , $L^{\mu \nu} \equiv k^{\mu}k^{\nu}/k^2$, and the right and left spinor projectors $ \omega_{\pm} \equiv \frac{1 \pm \gamma_5}{2} $, it is possible to derive from the SM renormalised Lagrangian the following vertex functions:
\begin{align}
    \Gamma^{\hat{A}\hat{A}}_{\mu \nu} \left(k, -k\right) = &
     -i T_{\mu\nu} \left[ \Sigma^{\hat{A}\hat{A}}_T(k^2) + k^2 Z_{AA} + \left(k^2-\mu_{Z}^2-\delta \mu_{Z}^2\right) \left(Z_{ZA}^\frac12 - 1\right)^2  \right] \nonumber\\
    &
    -i L_{\mu\nu} \left[ \Sigma^{\hat{A}\hat{A}}_L(k^2) -   \left(\mu_{Z}^2+\delta \mu_{Z}^2\right) \left(Z_{ZA}^\frac12 - 1\right)^2  \right]  \ ,
\\
    \Gamma^{\hat{Z}\hat{Z}}_{\mu \nu} \left(k, -k\right) = &
    -i T_{\mu\nu} \left[ \Sigma^{\hat{Z}\hat{Z}}_T(k^2) +  \left(k^2-\mu_Z^2-\delta \mu_Z^2\right) Z_{ZZ} + k^2 \left(Z_{AZ}^\frac12 - 1\right)^2 \right] \nonumber\\
    &
    -i L_{\mu\nu} \left[ \Sigma^{\hat{Z}\hat{Z}}_L(k^2) -\left(\mu_Z^2+\delta \mu_Z^2\right) Z_{ZZ}  \right]  \ ,
\\
    \Gamma^{\hat{A}\hat{Z}}_{\mu \nu} \left(k, -k\right) = &\ \Gamma^{\hat{Z}\hat{A}}_{\mu \nu} \left(k, -k\right) = \nonumber \\
    &
    -i T_{\mu\nu} \Big[ \Sigma^{\hat{A}\hat{Z}}_T(k^2) + k^2 Z_{AA}^\frac12 (Z_{AZ}^\frac12-1)\nonumber
    \\
    &\phantom{-i T_{\mu\nu}\Big[}
     + \left(k^2-\mu_{Z}^2-\delta \mu_{Z}^2\right) Z_{ZZ}^\frac12 \left(Z_{ZA}^\frac12-1\right)\Big] 
    \nonumber\\
    &
    -i L_{\mu\nu} \left[ \Sigma^{\hat{A}\hat{Z}}_L(k^2) - \left(\mu_{Z}^2+\delta \mu_{Z}^2\right) Z_{ZZ}^\frac12 \left(Z_{ZA}^\frac12-1\right) \right]  \ ,
\\
    \Gamma^{\hat{W}\hat{W}}_{\mu \nu} (k, -k) =& -i T_{\mu\nu} \left[ \Sigma^{\hat{W}\hat{W}}_T(k^2) + \left(k^2-\mu_W^2-\delta \mu_W^2\right) Z_{W} \right] \nonumber\\
    &
    -i L_{\mu\nu} \left[ \Sigma^{\hat{W}\hat{W}}_L(k^2) - \left(\mu_W^2+\delta \mu_W^2\right) Z_{W} \right]  \ ,
\\
    \Gamma^{\hat{H}\hat{H}} (k, -k) = &\ i \left[ \Sigma^{\hat{H}\hat{H}}(k^2) + \left(k^2-\mu_H^2-\delta \mu_H^2\right) Z_{H} \right] \ ,
\\
    \Gamma^{\hat{\chi}\hat{\chi}} (k, -k) =& \ i \left[ \Sigma^{\hat{\chi}\hat{\chi}}(k^2) + k^2 Z_{\chi} \right] \ ,
\\
    \Gamma^{\chi\chi} (k, -k) =& \ i \left[ \Sigma^{\chi\chi}(k^2) + (k^2-\mu_{\chi}^2-\delta \mu_{\chi}^2) Z_{\chi} \right] \ ,
\\
\Gamma^{\hat{\phi}^+\hat{\phi}^-} (k, -k) = &\ i \left[ \Sigma^{\hat{\phi}^+\hat{\phi}^-}(k^2) + k^2 Z_{\phi} \right] \ ,
\\
    \Gamma^{\phi^+ \phi^-} (k, -k) = &\ i \left[ \Sigma^{\phi^+ \phi^-}(k^2) + \left(k^2-\mu_{\phi}^2-\delta \mu_{\phi}^2\right) Z_{\phi} \right] \ ,
\\
    \Gamma^{\hat{A}\hat{H}}_{\mu} (k, -k) = &\ i k_{\mu} \Sigma^{\hat{A}\hat{H}}(k^2) \ ,
\\
    \Gamma^{\hat{A}\hat{\chi}}_{\mu} (k, -k) = &\ i k_{\mu} \Sigma^{\hat{A}\hat{\chi}}(k^2) \ ,
\\
    \Gamma^{\hat{Z}\hat{\chi}}_{\mu} (k, -k) = &\ i k_{\mu} \left[ \Sigma^{\hat{Z}\hat{\chi}}(k^2) + i \left(\mu_Z^2+\delta \mu_Z^2\right)^\frac12 Z_{ZZ}^\frac12 Z_\chi^\frac12 \right] \ ,
\\
    \Gamma^{\hat{W}^{\pm}\hat{\phi}^{\mp}}_{\mu} (k, -k) =& \ i k_{\mu} \left[ \Sigma^{\hat{W}^{\pm}\hat{\phi}^{\mp}}(k^2) \pm \left(\mu_W^2+\delta \mu_W^2\right)^\frac12 Z_{W}^\frac12 Z_\phi^\frac12 \right] \ ,
\\
    \Gamma^{f\bar{f}} (k, -k) = &\ i \slashed{k} \omega_{-} \left[ \Sigma^{f\bar{f}}_L (k^2) +{Z^f_L}^\frac12 {Z^{\bar{f}}_L}^\frac12 \right] + i \slashed{k} \omega_{+} \left[ \Sigma^{f\bar{f}}_R (k^2) + {Z^f_R}^\frac12 {Z^{\bar{f}}_R}^\frac12 \right]  \nonumber\\ 
    &
    -i \omega_{-} \left[ \Sigma^{f\bar{f}}_\ell (k^2) + \left(m_f+\delta m_f\right) {Z^f_L}^\frac12  {Z^{\bar{f}}_R}^\frac12  \right] \nonumber\\
    & 
    -i \omega_{+} \left[ \Sigma^{f\bar{f}}_r (k^2) + \left(m_f+\delta m_f\right) {Z^{\bar{f}}_{L}}^\frac12  {Z^f_{R}}^\frac12  \right]  \,,
\\
    \Gamma^{u_A\bar{u}_A} (k, -k) = &\ i \left[\Sigma^{u_A\bar{u}_A}(k^2) + k^2 \right] 
\\
    \Gamma^{u_Z\bar{u}_Z} (k, -k) = &\ i \left[\Sigma^{u_Z\bar{u}_Z}(k^2) + (k^2 - \mu_{u_Z}^2 - \delta \mu_{u_Z}^2) \right] 
\\
    \Gamma^{u_\pm\bar{u}_\pm} (k, -k) = & \ i \left[\Sigma^{u_\pm\bar{u}_\pm}(k^2) + (k^2 - \mu_{u_\pm}^2 - \delta \mu_{u_\pm}^2) \right] 
\\
    \Gamma^{\hat{H}} =& \ i \left[ T^{\hat{H}} + \left( t +\delta t \right) \sqrt{Z_{H}} \right] \ .
\end{align}
We label with $\Sigma$ the 1PI self-energy corrections,
which do not include tadpole contributions, indicated with $T^{\hat{H}}$
in the expression of the linear Higgs vertex function $\Gamma^{\hat{H}}$.
We discuss our tadpole convention in Section \ref{sec:Tadpole convention and Goldstone counterterms}. 
Looking at the pseudo-Goldstone bosons sector, the quantum fields two-point functions contain a  mass term $\mu_\chi^2 =\xi \mu_Z^2$ and $\mu_\phi^2=\xi \mu_W^2$, consistently renormalised, in agreement with the gauge-fixing choice of \cite{Denner:1994xt,Denner:2019vbn}.
The mass term of the $(\hat{\chi}, \hat{\phi}^{\pm})$ pseudo-Goldstone fields emerges from the introduction in the Lagrangian of an additional and independent gauge fixing for the background fields. In our case, we adopt the conventional $R_\xi$ gauge fixing for the background fields and choose the 't Hooft-Feynman gauge with $\xi=1$. The ghost two-point vertex functions contain mass terms $\mu^2_{u_Z} = \xi \mu_Z^2$ and $\mu^2_{u_\pm} = \xi \mu_W^2$, consistently with the Faddeev-Popov Langrangian term presented in \cite{Denner:1994xt,Denner:2019vbn}, to be renormalised with the associated gauge boson mass counterterms.

The CKM matrix has been set to the identity. For this reason the two-point vertex function $\Gamma^{f \bar{f}}$ describes only the diagonal self-energies between external fermions belonging to the same family. A complete treatment at first order that considers also the mixing matrix can be found in \cite{Denner:2019vbn}.

The on-shell renormalisation conditions are expressed exactly, to all orders in pertubation theory, by specific requests on the vertex functions and on their residues at specific kinematical points. The request that the propagator has a single pole at the physical mass is expressed by the vanishing of the vertex function at that specific virtuality. The probabilistic interpretation of the fields is preserved if the residue of the propagator is set equal to 1. 
\begin{align} 
\Gamma^{\hat{V}\hat{V}}_{\mu \nu} (k) \ \epsilon^{\nu}(k)\bigg|_{k^2 = \mu_V^2}  = &\;0 \,,
&
\lim_{k^2\to \mu_V^2}\frac{1}{k^2-\mu_V^2}\Gamma^{\hat{V}\hat{V}}_{\mu \nu} (k) \ \epsilon^{\nu}(k) =& - i \epsilon_{\mu}(k)  \,,
\nonumber\\
\Gamma^{\hat{A}\hat{Z}}_{\mu \nu} (k) \ \epsilon^{\nu}(k)\bigg|_{k^2 = 0}  = &\;0 \,,
&
\Gamma^{\hat{A}\hat{Z}}_{\mu \nu} (k) \ \epsilon^{\nu}(k)\bigg|_{k^2 = \mu_Z^2} =&\;0\,,
\nonumber\\
\Gamma^{\hat{H}\hat{H}}(k)|_{k^2 = \mu_H^2}  =&\; 0 \,,
&
\lim_{k^2\to \mu_H^2}\frac{1}{k^2-\mu_H^2}\Gamma^{\hat{H}\hat{H}}(k) =& \; i   \, ,
\nonumber\\
\Gamma^{f\bar{f}}(k) u_{f}(k)|_{k^2 = m_{f}^{2}} = &\;0 \, ,
&
\lim_{k^{2} \to m_{f}^{2}} \frac{\slashed{k}+m_{f}}{k^{2}-m_{f}^{2}}
\Gamma^{f \bar{f}} (k) u_{f}(k) =& \; iu_f(k)  \,,
\nonumber\\
\bar{u}_{f}(k)\Gamma^{f\bar{f}}(k)|_{k^2 = m_{f}^{2}} = & \; 0 \,,
&
\lim_{k^{2} \to m_{f}^{2}} \bar{u}_{f}(k) \Gamma^{f\bar{f}} ({k})\frac{\slashed{k}+m_{f}}{{k}^{2}-m_{f}^{2}} =&
\; i\bar u_f(k)  \,,
\end{align}
where $\varepsilon(k)$, $u(k)$ and $\bar{u}(k)$ are respectively the polarisation vectors and spinors of the external fields, $V$ indicates a generic gauge boson $A$, $Z$ or $W$ and $\mu_A=0$. In the CMS the pole of the propagator, where we impose the renormalisation conditions, is located in the complex plane. We discuss in Section \ref{sec:results} the implications of the presence of complex-valued renormalised parameters in the amplitude.
\subsection{On-shell counterterms }
\label{sec:onshellcts}
We expand the vertex functions to first order in the coupling constant and derive the expressions for the counterterms at this same order.
Let us focus on all the mass counterterms and on the bosonic WF counterterms.
\begin{align} 
\delta \mu^{2 (1)}_{Z} =& \Sigma_{T}^{\hat{Z}\hat{Z}(1)}(\mu^{2}_{Z}) \,,
\quad\quad
&
\delta \mu^{2 (1)}_{W} = &\Sigma_{T}^{\hat{W}\hat{W}(1)}(\mu^{2}_{W}) \,,
&&
\delta \mu^{2 (1)}_{H} = \Sigma_{T}^{\hat{H}\hat{H}(1)}(\mu^{2}_{H}) \,,
\nonumber\\
\delta m_{f}^{(1)} = &\mathrlap{\dfrac{m_{f}}{2}(\Sigma^{f\bar{f}(1)}_{L}(m_{f}^{2})+\Sigma^{f\bar{f}(1)}_{R}(m_{f}^{2})
+2\Sigma^{f\bar{f}(1)}_{S}({m_{f}}^{2})) \,,
} 
\nonumber\\
\delta Z^{(1)}_{AA} =& - \dfrac{\partial\Sigma_{T}^{\hat{A}\hat{A}(1)}(k^{2})}{\partial k^{2}} \bigg|_{k^{2}=0}  \ ,
& 
\delta Z^{(1)}_{ZZ} =& - \dfrac{\partial\Sigma_{T}^{\hat{Z}\hat{Z}(1)}(k^{2})}{\partial k^{2}} \bigg|_{k^{2}=\mu^{2}_{Z}} \,,
\nonumber\\
\delta Z^{(1)}_{AZ} =& - 2 \ \dfrac{\partial\Sigma_{T}^{\hat{A}\hat{Z}(1)}(\mu_Z^{2})}{\mu_Z^2} \,,
& 
\delta Z^{(1)}_{ZA} =& 2 \ \dfrac{\partial\Sigma_{T}^{\hat{Z}\hat{A}(1)}(0)}{\mu_Z^2} \,,
\nonumber\\
\delta Z^{(1)}_{H} =& - \dfrac{\partial\Sigma_{T}^{\hat{H}\hat{H}(1)}(k^{2})}{\partial k^{2}} \bigg|_{k^{2}=\mu^{2}_{H}} \,,
& 
\delta Z^{(1)}_{W} =& - \dfrac{\partial\Sigma_{T}^{\hat{W}\hat{W}(1)}(k^{2})}{\partial k^{2}} \bigg|_{k^{2}=\mu^{2}_{W}} \,,
\end{align}
 where we have  exploited the relation
\begin{equation}
    \Sigma^{f\bar{f}}_r (k^2) = \Sigma^{f\bar{f}}_s (k^2) = \frac{1}{2} \Sigma^{f\bar{f}}_S (k^2) \ , 
\end{equation}
that has been explicitly verified at first and second perturbative order.

\begin{figure}
\centering
\includegraphics[width=0.9\textwidth, trim= 0cm 3cm 0cm 5cm]{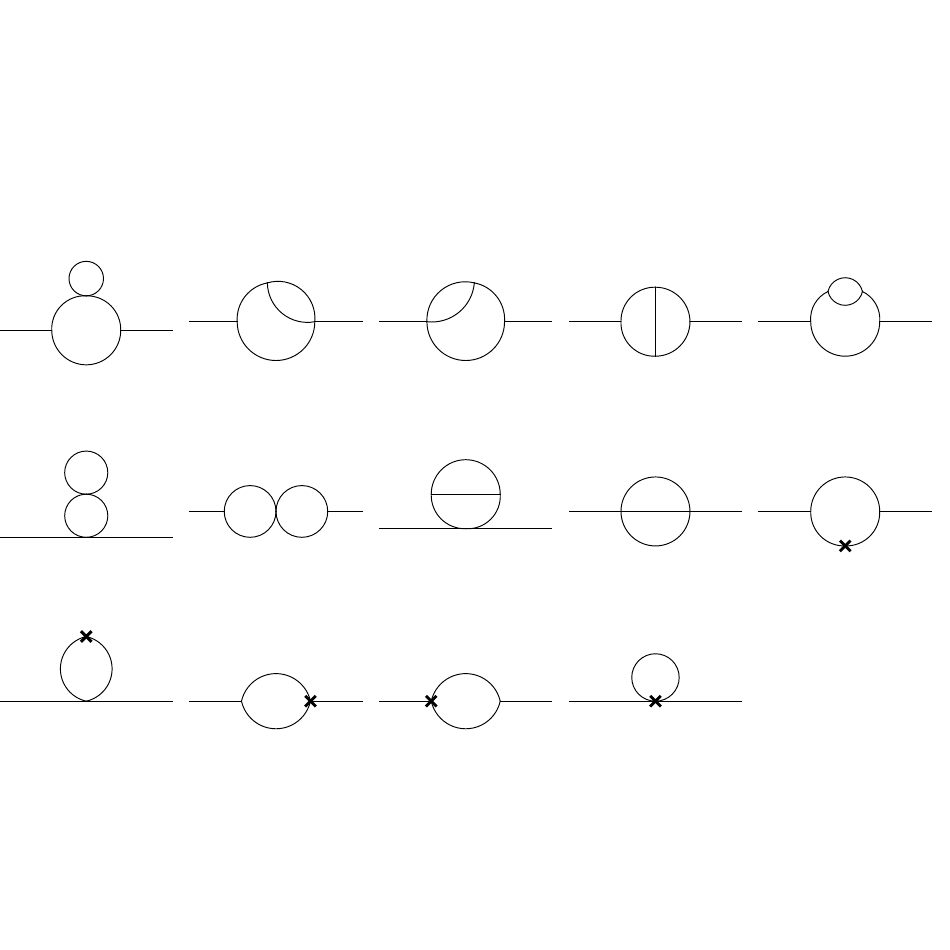}
\caption{\label{fig:Sigma_contributions} 1PI topologies that contribute to the definition of $\mathcal{O}(\alpha^2)$ sub-renormalised self-energies $\Sigma^{(2)}$.}
\end{figure}

Applying the renormalisation conditions to the second order expanded vertex functions and simplifying the result with the first-order expressions, we derive the second-order counterterm definitions:
\begin{align} \label{eq:CTs2ndorder}
\delta Z^{(2)}_{AA} =& - \dfrac{\partial\Sigma_{T}^{\hat{A}\hat{A}(2)}(k^{2})}{\partial k^{2}} \bigg|_{k^{2}=0}  \ ,
\nonumber\\
\delta Z^{(2)}_{ZZ} =& - \dfrac{\partial\Sigma_{T}^{\hat{Z}\hat{Z}(2)}(k^{2})}{\partial k^{2}} \bigg|_{k^{2}=\mu^{2}_{Z}} - \dfrac{1}{4} (\delta Z^{(1)}_{AZ})^2  \ ,
\nonumber\\
\delta \mu^{2 (2)}_{Z} =& \Sigma_{T}^{\hat{Z}\hat{Z}(2)}(\mu^{2}_{Z}) - \delta \mu^{2 (1)}_{Z} \delta Z^{(1)}_{Z} +\dfrac{1}{4} \mu^2_Z \left(\delta Z^{(1)}_{AZ}\right)^2 \ ,
\nonumber\\
\delta \mu^{2 (2)}_{W} =& \Sigma_{T}^{\hat{W}\hat{W}(2)}(\mu^{2}_{W}) - \delta \mu^{2 (1)}_{W} \delta Z^{(1)}_{W} \ ,
\nonumber\\
\delta Z^{(2)}_{W} = & - \dfrac{\partial\Sigma_{T}^{\hat{W}\hat{W}(2)}(k^{2})}{\partial k^{2}} \bigg|_{k^{2}=\mu^{2}_{W}}  \ ,
\nonumber\\
\delta Z^{(2)}_{AZ} =& - 2 \ \dfrac{\partial\Sigma_{T}^{\hat{A}\hat{Z}(2)}(\mu_Z^{2})}{\mu_Z^2} -\dfrac{1}{2} \delta Z^{(1)}_{AA} \delta Z^{(1)}_{AZ} +\dfrac{1}{4} \left(\delta Z^{(1)}_{AZ}\right)^2 \ ,
\nonumber\\
\delta Z^{(2)}_{ZA} =& 2 \ \dfrac{\partial\Sigma_{T}^{\hat{Z}\hat{A}(2)}(0)}{\mu_Z^2} -\dfrac{1}{2} \delta Z^{(1)}_{ZZ} \delta Z^{(1)}_{ZA} +\dfrac{1}{4} (\delta Z^{(1)}_{ZA})^2 - \dfrac{1}{\mu^2_Z} \delta \mu^{2 (1)}_{Z} \delta Z^{(1)}_{ZA} \ ,
\nonumber\\
\delta \mu^{2 (2)}_{H} = &\Sigma_{T}^{\hat{H}\hat{H}(2)}(\mu^{2}_{H}) - \delta \mu^{2 (1)}_{H} \delta Z^{(1)}_{H} \ ,
\nonumber\\
\delta Z^{(2)}_{H} =& - \dfrac{\partial\Sigma_{T}^{\hat{H}\hat{H}(2)}(k^{2})}{\partial k^{2}} \bigg|_{k^{2}=\mu^{2}_{H}}  \ ,
\nonumber\\
\delta m_{f}^{(2)} =& \dfrac{m_{f}}{2}\bigg[\Sigma^{f\bar{f}(2)}_{L}(m_{f}^{2})+\Sigma^{f\bar{f}(2)}_{R}(m_{f}^{2})
+2\Sigma^{f\bar{f}(2)}_{S}({m_{f}}^{2}) - \dfrac{1}{2} \delta Z^{f(1)}_{L} \delta Z^{f(1)}_{R} \nonumber\\ 
& + \dfrac{1}{4} \left(\delta Z^{f(1)}_{L}\right)^2 + \dfrac{1}{4} \left(\delta Z^{f(1)}_{R}\right)^2 \bigg] - \dfrac{1}{2} \delta m_{f}^{(1)} \left( \delta Z^{f(1)}_{S} + \delta Z^{f(1)}_{R} + \Sigma^{f\bar{f}(1)}_{S}({m_{f}}^{2}) \right) \ .
\end{align}
Considering our $\mathcal O(\alpha^2)$ computation, the self-energies $\Sigma^{(2)}$ are defined to contain also the contributions that we refer to as \textit{sub-renormalisation} diagrams, namely the insertion of a first order counterterm (coupling or wavefunction) in a 1PI one-loop diagram, as shown in Fig.~\ref{fig:Sigma_contributions}.
An additional element to take into account when performing a two-loop computation is the presence in the second order counterterm definitions of products between first order counterterms, originating from the expansion up to $\mathcal{O}(\alpha^2)$ of the renormalised $\Gamma$ functions.

Let us now focus on the case of the fermionic WF counterterms.
The fermion and antifermion fields may in principle be renormalised with different WF constants, namely $Z^{f}_{L/R}$ and $Z^{\bar{f}}_{L/R}$, spoiling at the renormalised level the bare relation $\bar{f}_0= f^{\dagger}_0 \gamma_0$, which ensures the hermiticity of the Lagrangian. 
On the other hand, the first and second order renormalisation conditions  constrain  the sum $\delta Z^{f}_{L/R}+\delta Z^{\bar{f}}_{L/R}$. Since the difference $\delta Z^{f}_{L/R}-\delta Z^{\bar{f}}_{L/R}$ introduces only a relative global phase between the renormalised fields $f$ and $\bar{f}$, as indicated in \cite{Denner:2019vbn}, one can conclude that we have the freedom of fixing at every order:
\begin{equation}
    \delta Z^f_{L/R} = \delta Z^{\bar{f}}_{L/R} \ .
\end{equation}
Adding this simplification to the renormalisation conditions we obtain: 
\begin{align}
&\delta Z^{f(1)}_{L} = -\Sigma^{f\bar{f}(1)}_{L}(m_{f}^{2}) - m_{f}^{2}\dfrac{\partial}{\partial k^{2}} \left(\Sigma^{f\bar{f}(1)}_{L}(k^{2}) +\Sigma^{f\bar{f}(1)}_{R}(k^{2}) + 2\Sigma^{f\bar{f}(1)}_{S}(k^{2})\right)\bigg|_{k^{2}=m_{f}^{2}} \ ,
\nonumber\\
&\delta Z^{f(1)}_{R} = -\Sigma^{f\bar{f}(1)}_{R}(m_{f}^{2}) - m_{f}^{2}\dfrac{\partial}{\partial k^{2}} \left(\Sigma^{f\bar{f}(1)}_{L}(k^{2}) +\Sigma^{f\bar{f}(1)}_{R}(k^{2}) + 2\Sigma^{f\bar{f}(1)}_{S}(k^{2})\right)\bigg|_{k^{2}=m_{f}^{2}} \ .
\nonumber\\
&\delta Z^{f(2)}_{L} = -\Sigma^{f\bar{f}(2)}_{L}(m_{f}^{2}) + m_{f} \ \delta m_{f}^{(1)} \dfrac{\partial \Sigma^{f\bar{f}(1)}_{S}({m_{f}}^{2})}{\partial k^2}\bigg|_{k^{2}=m_{f}^{2}} \nonumber \\ 
& \phantom{\delta Z^{f(2)}_{L} = } - m_{f}^{2}\dfrac{\partial}{\partial k^{2}} \left(\Sigma^{f\bar{f}(2)}_{L}(k^{2}) +\Sigma^{f\bar{f}(2)}_{R}(k^{2}) + 2\Sigma^{f\bar{f}(2)}_{S}(k^{2})\right)\bigg|_{k^{2}=m_{f}^{2}} \ , 
\nonumber\\
&\delta Z^{f(2)}_{R} = -\Sigma^{f\bar{f}(2)}_{R}(m_{f}^{2}) + m_{f} \ \delta m_{f}^{(1)} \dfrac{\partial \Sigma^{f\bar{f}(1)}_{S}({m_{f}}^{2})}{\partial k^2}\bigg|_{k^{2}=m_{f}^{2}} \nonumber \\ 
&\phantom{\delta Z^{f(2)}_{R} =} - m_{f}^{2}\dfrac{\partial}{\partial k^{2}} \left(\Sigma^{f\bar{f}(2)}_{L}(k^{2}) +\Sigma^{f\bar{f}(2)}_{R}(k^{2}) + 2\Sigma^{f\bar{f}(2)}_{S}(k^{2})\right)\bigg|_{k^{2}=m_{f}^{2}} \  . \label{eq:deltaZf2}
\end{align}

The CMS leads to a dependence of the 
scattering amplitude
on the internal complex masses \mz \ and \mw, and in turn to a violation of the optical theorem, by terms of \oaaa \cite{Denner:2014zga}. The clear advantage of a uniform systematic usage of the complex masses in all the Green's functions is the explicit fulfillment of gauge invariance, but also the exact cancellation of the UV divergences. As detailed in \cite{Frederix:2018nkq}, if the final state particles are unstable, with a non-vanishing decay width,  but we insist in treating them as asymptotic states, then in the evaluation of their WF renormalisation constants we have to remove the absorbitive part. This subtraction can be achieved with the introduction of a suitable $\widetilde{\text{Re}}$ operator in Equations \ref{eq:deltaZf2}, in analogy to the one-loop example presented in \cite{Frederix:2018nkq}, and will be discussed in a separate publication.

\subsection{Renormalisation conditions following from the BFG Ward Identities} \label{sec:rencondWardId}
The $U(1)_{em}$ gauge invariance of the classical SM Lagrangian
translates into QED-like Ward identities that connect the 1PI vertex functions involving external background photon fields. The complete list of identities valid in the Background Field Gauge chosen in the present calculation can be found in \cite{Weiglein:1993hd, Denner:1994nn, Denner:2019vbn}. In the case of the bare 1PI gauge boson self-energies, we have relations connecting the 2-point functions with 2 external vector bosons, with 1 vector boson and 1 pseudo-Goldstone boson and with 2 Goldstone bosons:
\begin{align}
\Sigma_{L \ \text{(bare)}}^{\hat{A}\hat{A}}(k^2)=&\,0 \,, \nonumber\\
\Sigma_{L \ \text{(bare)}}^{\hat{A}\hat{Z}}(k^2)=&\,0  \,,\nonumber\\
\Sigma_{\text{(bare)}}^{\hat{A}\hat{\chi}}(k^2)=&\,0  \,,\nonumber\\
\Sigma_{\text{(bare)}}^{\hat{A}\hat{H}}(k^2)=&\,0  \,,\nonumber\\
\Sigma_{L \ \text{(bare)}}^{\hat{Z}\hat{Z}}(k^2) - i \mu_Z \Sigma_{\text{(bare)}}^{\hat{Z}\hat{\chi}}(k^2)=&\,0  \,,\nonumber\\
k^2 \Sigma_{\text{(bare)}}^{\hat{Z}\hat{\chi}}(k^2) - i \mu_Z \Sigma_{\text{(bare)}}^{\hat{\chi}\hat{\chi}}(k^2) + i \frac{e}{2 c_W s_W} T^{\hat{H}}_{\text{(bare)}}=&\,0  \,,\nonumber\\
\Sigma_{L \ \text{(bare)}}^{\hat{W}^{\pm}\hat{W}^{\mp}}(k^2) \mp \mu_W \Sigma_{\text{(bare)}}^{\hat{W}^\mp \hat{\phi}^\pm}(k^2)=&\,0 \,,\nonumber\\
k^2 \Sigma_{\text{(bare)}}^{\hat{W}^\mp \hat{\phi}^\pm}(k^2) \pm \mu_W \Sigma_{\text{(bare)}}^{\hat{\phi}^\mp\hat{\phi}^\pm}(k^2) \mp \frac{e}{2 s_W} T^{\hat{H}}_{\text{(bare)}}=&\,0 \ ,
\label{eq:wardid}
\end{align}
with $s_W = \sin \theta_W, c_W = \cos \theta_W$. For the analyticity of the associated two-point vertex functions at $k^2=0$, the first two conditions also imply:
\begin{align}
\Sigma_{T \ \text{(bare)}}^{\hat{A}\hat{A}}(0)&=0  \,,\nonumber \\
\Sigma_{T \ \text{(bare)}}^{\hat{A}\hat{Z}}(0)&=0 \,,
\end{align}
and in turn that
$\delta Z_{ZA}=0$  at all orders in perturbation theory.
We have verified these relations for all the unrenormalised gauge boson self-energies in two configurations, namely restricting the check to the fermionic subset that will be described in Section \ref{eq:amprenormalised} or including the complete set of two-loop EW corrections. The verification of these relations is a significant technical check of our computational workflow.

The Ward identities in Equation (\ref{eq:wardid}), together with those satisfied by the three-point functions, can be exploited, upon renormalisation, for an alternative definition of the WF counterterms. At second order, indeed, the expression of the gauge boson mass counterterms requires the knowledge of the corresponding first order WF renormalisation constant. The latter can be determined either by setting to one the residue of the propagator (unitarity condition), as described in Section \ref{sec:onshellcts}, or by imposing the validity of the second order Ward identities also at the renormalised level (gauge invariance condition).
At first order the two possibilities differ in the finite part.
For the sake of building a gauge invariant scattering amplitude, we choose the second alternative.

At the one-loop level, the Ward Identities satisfied by the two-point connected vertex functions lead, after renormalisation,   to the relations:
\begin{align} \label{eq:deltaZchi1}
    \delta Z_\chi^{(1)} = \delta Z_{ZZ}^{(1)} + \frac{\delta \mu_Z^{2 (1)}}{\mu_Z^2} \ ,\\
    \delta Z_\phi^{(1)} = \delta Z_{W}^{(1)} + \frac{\delta \mu_W^{2 (1)}}{\mu_W^2} \ ,
\end{align}
while the renormalised Ward Identities featuring three-point vertex functions lead to:
\begin{align}
    &\delta Z_{AA}^{(1)} = - 2 \delta Z_e^{(1)} \ , \\
    &\delta Z_{AZ}^{(1)} = - 2 \frac{\delta s_W^{2(1)}}{c_W s_W} \ , \\
    &\delta Z_{ZZ}^{(1)} = -2 \delta Z_e^{(1)} + \frac{c_W^2-s_W^2}{c_W^2} \frac{\delta s_W^{2 (1)}}{s_W^2} \ , \\
    &\delta Z_W^{(1)} = -2 \delta Z_e^{(1)} +  \frac{\delta s_W^{2 (1)}}{s_W^2} \ , \\
    &\delta Z_H^{(1)} = -2 \delta Z_e^{(1)} +  \frac{\delta s_W^{2 (1)}}{s_W^2} +  \frac{\delta \mu_W^{2 (1)}}{\mu_W^2} \ .
\end{align}
The condition on the photon wavefunction counterterm $\delta Z_{AA}^{(1)}$ is in perfect agreement with the all-order condition $Z_e=Z_{AA}^{-1/2}$ that reflects the universality of the electric charge, whose renormalisation receives non-vanishing contributions only from the photon vacuum polarisation.
The other relations directly imply:
\begin{align}
    \delta Z_H^{(1)} = \delta Z_{\phi}^{(1)} = \delta Z_\chi^{(1)} \ ,
\end{align}
representing the condition that at \oa all the components of the complex Higgs field doublet must be renormalised with the same wavefunction constant. In addition, one would also obtain for the renormalisation of the left-handed fermion fields:
\begin{equation} \label{eq:deltaZfL1}
    \delta Z_{f_u}^{L (1)} = \delta Z_{f_d}^{L (1)} \ ,
\end{equation}
constraining also the wavefunction constants of all the left-handed fermions belonging to the same doublet to be the same.

In general, Equations (\ref{eq:deltaZchi1}-\ref{eq:deltaZfL1}) spoil the probabilistic interpretation of the Higgs and gauge bosons renormalised propagators that is ensured in the on-shell scheme with the requirement of the associated pole residues fixed to one. Hence, if a gauge boson or Higgs field appears as an external particle in the computation of an S-matrix element, the residue of the associated renormalised propagator must be consistently brought to one at the end of the computation through an ad-hoc UV-finite renormalisation, as presented in \cite{Denner:1996gb, Denner:2019vbn}. This additional procedure is not required for the case of the neutral-current Drell-Yan process featuring purely fermionic external states, and its application for general two-loop processes is left to future publications.

\subsection{Electric charge renormalisation within the \texorpdfstring{$\overline{MS}$}{MS} scheme}

The BFG Ward Identity associated to the residual $U(1)_{em}$ gauge invariance of the background photon field ensures that the renormalisation of the electric charge follows entirely from the transverse photon-photon self-energy $\Sigma_{T}^{AA}(k^2) $\cite{Denner:1994xt,Degrassi:2003rw,Dittmaier:2021loa}, via the renormalisation condition on the $\gamma f f$ vertex that relates $Z_e$ and $Z_{AA}$:
\begin{equation} \label{dZedZAA relation}
    Z_e = \ \frac{1}{\sqrt{Z_{AA}}} \ \ \ \ \text{with}  \ \ \ \delta Z_{AA} = - \dfrac{\partial\Sigma_{T}^{AA}(q^{2})}{\partial k^{2}} \bigg|_{k^{2}=0} \; .\;
\end{equation}
We consider the electric charge renormalisation in the $\msbar$ scheme, for consistency with our subtraction term of the IR divergences, described in Section \ref{sec:infra}.
We make explicit the dependence on the renormalisation scale $\mu_R$ and on the regularisation 't Hooft scale $\mu$, and we factor out for each loop integration an overall factor $S_\epsilon=(4\pi e^{-\gamma_E})^\epsilon$:
\begin{align}
\delta Z_e^{\overline{MS}(1)}&= \ S_\epsilon \left( \frac{\mu^{2}}{\mu_R^{2}}\right)^\epsilon\left(-\frac{1}{2} \delta Z_{AA}^{OS(1)}\big|_\text{poles}\right) \; , \nonumber\\
\delta Z_e^{\overline{MS}(2)}&= \ S_\epsilon^2 \left( \frac{\mu^{2}}{\mu_R^{2}}\right)^{2\epsilon} \left(-\frac{1}{2} \delta Z_{AA}^{OS(2)}\big|_\text{poles}+\frac{3}{8} \left(\delta Z_{AA}^{OS(1)}\big|_\text{poles} \right)^2\right)  \; .
\end{align}
In our computation, both the UV renormalisation procedure and the IR subtraction of divergences have been realised fixing for convenience $\mu=\mu_R$.
The presence in the coefficient of the UV pole of the masses of the gauge bosons implies that the electric charge counterterm in the $\msbar$ scheme is complex valued; the same comment applies a fortiori in the on-shell renormalisation scheme. 

\subsection{Tadpole convention and Goldstone counterterms} \label{sec:Tadpole convention and Goldstone counterterms}
Expanding the bare Lagrangian scalar potential $V(\hat\Phi)=- \mu_0^2 (\hat\Phi_0^{\dagger} \hat\Phi_0) + \lambda_0(\hat\Phi^{\dagger}_0 \hat\Phi_0)^2/4$, with $\hat\Phi_0$ representing the background Higgs doublet $\hat\Phi=(\hat\phi^+_0,v_0+\hat H_0+i\hat\chi_0)^T$, we obtain a linear term $t_0\hat{H}(x)$ with $t_0$ being a function of the bare fundamental parameters:
\begin{equation}
    t_0=v_0 \left(\mu_0^2-\lambda_0 \frac{v_0^2}{4}\right) .
\end{equation}
The value of the bare Higgs field vacuum expectation value (VEV) can be fixed by requiring the minimisation of the Higgs scalar potential, solving $\partial V(|\hat{\Phi}|^2)/\partial |\hat{\Phi}|^2 = 0$. The result is
\begin{equation}
    v_0^2=\frac{4\mu_0^2}{\lambda_0} \ ,
\end{equation}
immediately implying two main observations valid at tree-level: 
\begin{itemize}
    \item $t_0=0$, representing the fact that no tree-level Higgs interactions  with the vacuum are permitted;
    \item $\mu_0^2$ is directly related to the Higgs mass through $\mu_{H,0}^2 = - \mu_0^2 + \frac{3}{4} \lambda_0 v_0^2 = 2 \mu_0^2$.
\end{itemize}
\begin{figure}
\begin{minipage}{0.25\textwidth}
\centering
\includegraphics[width=0.90\textwidth]{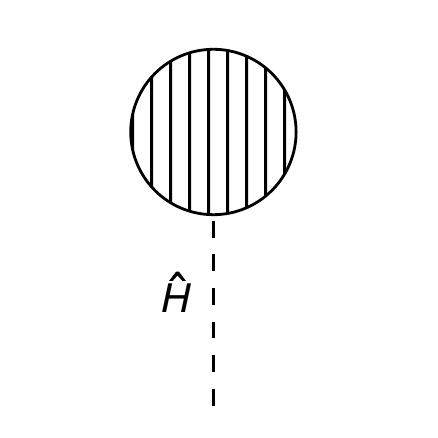} \subcaption{}
\end{minipage}
\begin{minipage}{0.25\textwidth}
\centering
\includegraphics[width=0.90\textwidth]{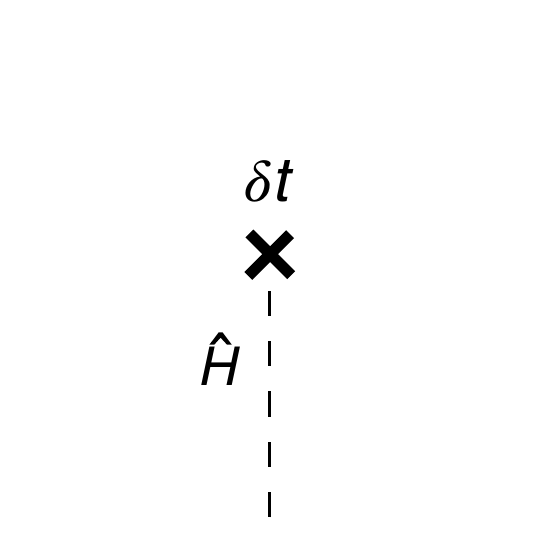}
\subcaption{}
\end{minipage}
\centering
\caption{\label{fig:TH and deltat} Diagrammatic representation of the $T^{\hat{H}}$ tadpole topology for the background Higgs field (a) and associated $\delta t$ counterterm insertion (b).}
\end{figure}

The quantum corrections to the Higgs potential lead to an effective potential of the form $V_{eff}(\hat{\Phi}) = V(\hat{\Phi}) + \Delta V$, resulting in an higher-order Higgs VEV to be modified with respect to the tree-level one. Nevertheless, it is possible to exploit the invariance of the SM theory with respect to a possible shift in the bare background Higgs field $\hat{H}_0$ or in the bare parameters $v_0$, $\lambda_0$ and $\mu_0$, and reorganise this kind of higher order corrections. The arbitrariness of this procedure leads to the different possible choices of tadpole schemes, such as the Fleischer-Jegerlehner tadpole scheme (FJTS) \cite{Fleischer:1980ub} and the $\beta_h$ tadpole scheme \cite{Actis:2006ra}. In our case, we decided to fix the renormalised value of the Higgs VEV to obey its tree-level definition, following the Parameter-Renormalised Tadpole Scheme (PRTS) described in \cite{Sirlin:1980nh, Sirlin:1985ux, Denner:1994xt,Denner:2019vbn}:
\begin{equation} \label{RenormHiggsvev}
    v^2 = \frac{4 \mu^2}{\lambda}  \ \ \ \implies \ \ \ t=0 \ , 
\end{equation}
by shifting the bare parameter $t_0$ and introducing at the same time a $\delta t$ counterterm ($t_0=t+\delta t$) satisfying the following reparametrisation condition:
\begin{equation} 
    \Gamma^{\hat{H}} = T^{\hat{H}} + (t+\delta t) \sqrt{Z_H} = T^{\hat{H}} + \delta t \sqrt{Z_H} = 0 \ .
\end{equation}
The final effect is the enforcement at the renormalised level of the diagrammatic cancellation between the $\delta t$ counterterm contribution and the higher order Higgs tadpole diagrams, indicated with a striped bubble in Fig.\ref{fig:TH and deltat}. The first and second order tadpole counterterms read:
\begin{align}
    &\delta t^{(1)} = - T^{H(1)} \ , \\
    &\delta t^{(2)} = - T^{H(2)} + \frac{1}{2} T^{H(1)} \delta Z_H^{(1)} \ .
\end{align}
This choice is very useful in both the computation of the virtual amplitude and the evaluation of the
UV counterterms. The introduction of $\delta t$ propagates the $T^{\hat{H}}$ contribution in the system of equations that connects the input variables counterterms with those associated to the fundamental Lagrangian parameters:
\begin{align}
\begin{cases}
    \; t + \delta t = - T^{\hat{H}} = \left( v + \delta v \right) \left( \mu^2+\delta\mu^2-\dfrac{1}{4}\left(\lambda + \delta \lambda \right) \left(v + \delta v\right)^2 \right)
    \\ \\[-15pt]
   \;  \delta e = \dfrac{\left( g + \delta g \right) \left( g' + \delta g' \right)}{\sqrt{\left( g + \delta g \right)^2+\left( g' + \delta g' \right)^2}}
    \\ \\[-15pt]
    \; \delta \mu_H^2 = \dfrac{3}{4} \left( \lambda + \delta \lambda \right)\left(v + \delta v\right)^2 - \mu^2 -\delta \mu^2
    \\ \\[-15pt]
    \; \delta \mu_W^{2} = \dfrac{1}{4} \left(v + \delta v\right)^2 \left(g + \delta g\right)^2
    \\ \\[-15pt]
    \; \delta \mu_Z^{2} = \dfrac{1}{4} \left(v + \delta v\right)^2 \left( \left(g + \delta g\right)^2 + \left(g' + \delta g'\right)^2 \right)
\end{cases}
\end{align}

A first consequence of the tadpole diagrams reorganisation affects the $\delta \mu^2$ counterterm defined from $\mu_0^2 = \mu^2 + \delta \mu^2$. While Equation \ref{RenormHiggsvev} implies that also at the renormalised level
\begin{equation}
    \mu_H^2 = 2 \mu^2 = \lambda \frac{v^2}{2} \ ,
\end{equation}
the expression of $\delta \mu^2$ does not  trivially follow from the bare relation, receiving a shift from the tadpole contributions already at the first order:
\begin{align}
    &\delta \mu^{2 \ (1)} = \frac{1}{2} \left( \delta \mu_H^{2 \ (1)} - \frac{3 T^{H (1)}}{v} \right)
\end{align}
The same logic applies to the Goldstone boson counterterms derivation. When isolating in the scalar potential $V(\Phi+\hat{\Phi})$ the quadratic terms in the background and quantum fields $(\hat{\phi}^{\pm},\hat{\chi},\phi^\pm, \chi)$, additional mass terms appear showing a proportionality with $t_0$:
\begin{align} 
    -\frac{t_0}{v_0} \ \hat{\phi}^+ \hat{\phi}^-\,,  \quad -\frac{t_0}{v_0} \ \frac{1}{2} \hat{\chi}^2(x)\,, \quad -\frac{t_0}{v_0} \ \phi^+ \phi^-\,,  \quad -\frac{t_0}{v_0} \ \frac{1}{2} \chi^2(x)\,.
\end{align}
As a result, a shift of the mass counterterms with respect to those emerging purely from the gauge fixing part of the Lagrangian must be considered, giving for the quantum fields
\begin{align}
    \delta \mu_{\phi}^{2(1)} =\ & \xi \delta \mu_W^{2(1)} + \frac{T^{\hat{H}(1)}}{v} \,,\nonumber\\
    \delta \mu_{\chi}^{2(1)} = \ &\xi \delta \mu_Z^{2(1)} + \frac{T^{\hat{H}(1)}}{v} \,,\nonumber\\
    \delta \mu_{\phi}^{2(2)} =\  &\xi \delta \mu_W^{2(2)} + \frac{T^{\hat{H}(2)}}{v} - \frac{\delta v^{(1)} T^{\hat{H}(1)}}{v^2} - \frac{\delta Z_H^{(1)} T^{\hat{H}(1)}}{2v} \,,\nonumber\\
    \delta \mu_{\chi}^{2(2)} = \ &\xi \delta \mu_Z^{2(2)} + \frac{T^{\hat{H}(2)}}{v}  - \frac{\delta v^{(1)} T^{\hat{H}(1)}}{v^2} - \frac{\delta Z_H^{(1)} T^{\hat{H}(1)}}{2v}\,,
\end{align}
as well as $ \delta \mu_{\hat{\phi}}^{2} = \delta \mu_{\phi}^{2} - \xi \delta \mu_W^2$ and $ \delta \mu_{\hat{\chi}}^{2} =\delta \mu_{\chi}^{2} - \xi \delta \mu_Z^2$ in the case of background fields.

\subsection{Semi-analytical evaluation of the SM renormalisation constants}
\label{sec:OS_limits}
The renormalisation constants at first and second order have been evaluated in dimensional regularisation as a series expansions in $\epsilon=(4-d)/2$, and collected in a dedicated library. In order to perform consistent two-loop amplitude computations, the first order counterterms have been evaluated up to $\mathcal{O}(\epsilon^2)$, while the second order ones up to $\mathcal{O}(\epsilon^0)$.
The evaluation of the WF renormalisation constants in the on-shell scheme requires a correct treatment of the on-shell limits 
$k^2\to m^2$
for the functions involved in the calculation.
Such limits in general do not commute with the $\epsilon$ expansion, and have to be performed first to obtain the correct representation of the IR poles in dimensional regularisation.
The evaluation of the derivative of the self-energy functions requires the computation of the derivatives of the relevant scalar Feynman integrals. The latter can be obtained, keeping the exact dependence on $\epsilon$, with standard algorithms \cite{Lee:2013mka},
and the coefficients which appear in the equations are free of spurious poles $(k^2-m^2)^{-\alpha}$.
At this point, we pay attention to use integration-by-part relations computed with on-shell kinematics, to reduce the scalar integrals to a minimal set of Master Integrals. This special choice again
avoids the appearance of spurious poles.
The value of the Master Integrals in the on-shell point $k^2=m^2$ can be evaluated with \amflow\, if $m\neq 0$
or with the analytical results of Ref. \cite{Davydychev:1992mt}
for $m=0$. Additional details on the evaluation of the Master Integrals are presented in Appendix \ref{app:semianev}. 

Our choice of performing the computation in the Complex Mass Scheme leads to a non-trivial interplay between the Feynman prescription and the particle decay width. For a self-energy $\Sigma(k^2)$, the evaluation of the mass and wavefunction renormalisation constants requires  the analytic continuation of $k^2$ from the upper half of the $k^2$ complex plane (Feynman prescription for the Real Mass Scheme) to the lower half plane (due to $\mu_V^2=m_V^2-i\Gamma_V m_V$, position of the pole of the propagator), leading to an unphysical discontinuity of the function when the real axis is crossed. We can equivalently say that the limit of vanishing decay width is discontinuous. The correct prescription for the evaluation of these counterterms is thus on the second Riemann sheet of the $k^2$ complex surface~\cite{Passarino:2010qk,Denner:2019vbn}.
The analytic continuation algorithm implemented in the codes \seasyde~\cite{Armadillo:2022ugh} or \amflow~\cite{Liu:2022chg} allows a straightforward evaluation of the counterterms on the second Riemann sheet.

\section{Infrared subtraction}
\label{sec:infra}
Considering the case of fermionic EW corrections up to second order, we define an UV-renormalised and IR-subtracted finite remainder as follows:
\begin{align}
  | \M^{(0,1),fin}_{\rm fer} \rangle &= 
  | \M^{(0,1)}_{\rm fer} \rangle - \;\I^{(0,1)}_{\rm fer} | \M^{(0,0)} \rangle
~=~| \M^{(0,1)}_{\rm fer} \rangle \label{eq:subtractedone}\,,\\
  | \M^{(0,2),fin}_{\rm fer} \rangle &= 
  | \M^{(0,2)}_{\rm fer} \rangle - \; \I^{(0,2)}_{\rm fer} | \M^{(0,0)} \rangle
                      -  \;{\I}^{(0,1)} | \M^{(0,1),fin}_{\rm fer} \rangle\,.
 \label{eq:subtracted}
\end{align} 
We refer to \cite{Armadillo:2025mfx} for the complete discussion on the IR subtraction term valid at second order in the EW SM and the general definitions of the functions $\I^{(0,n)}$, which can be obtained from the abelianisation of the QCD soft and cusp anomalous dimensions, while in the following we focus on the subtraction of IR poles in the specific case of the fermionic contribution to the amplitude $| \M^{(0,n),fin}_{\rm fer} \rangle$.
The subtraction term is a proxy of the real-emission matrix elements integrated over the soft/collinear phase space of the emitted partons. In the case of the fermionic corrections, we need to consider three contributions: the emission of one photon receiving a fermionic correction in the coupling,  the tree-level emission of a fermion pair, the tree-level emission of a photon which factors with respect to a hard matrix element with a hard fermionic correction.
At one-loop the only contributing fermionic diagrams are  gauge-boson self-energies inserted in the tree level propagators, and are not IR divergent; consistently, the subtraction operator $\I^{(0,1)}_{\rm fer}=0$ because at first order the emission of a photon is not of fermionic kind.
The infrared structure of the virtual amplitude is thus encapsulated in the subtraction operators $\I^{(0,1)}$ and  $\I^{(0,2)}_{\rm fer}$, defined as
\begin{align}
\label{eq:I01}
\frac 1{S_\epsilon}\,\I^{(0,1)} =& \
\frac{\Gamma'_0}{4\epsilon^2}+\frac{\Gamma_0}{2\epsilon}
\,,\\
\label{eq:I02}
\frac 1{S_\epsilon^{2}}\,\I^{(0,2)}_{\rm fer} =& \ -\frac{3\Gamma'_0}{16\epsilon^3}\beta_0-\frac{\Gamma_0}{4\epsilon^2}\beta_0+\frac{\Gamma'_{1}}{16\epsilon^2}+\frac{\Gamma_{1fer}}{4\epsilon}
-\frac23 \sum_{k}^{\text{massive}} n_k q_k^2 
\\&\nonumber
\times
\left\{\frac1{2\epsilon^2} \Gamma'_0\log\left(\frac{\mu_R^2}{m_k^2}\right) + \frac1{4\epsilon}\left[ \Gamma'_0\left(\log^2\left(\frac{\mu_R^2}{m_k^2}\right)+\frac{\pi^2}6 \right)+4\Gamma_0 \log\left(\frac{\mu_R^2}{m_k^2}\right)\right]\right\}\;,
\end{align}
where $S_\epsilon=(4\pi e^{-\gamma_E})^\epsilon$ and $\mu_R$ is the renormalisation scale. With respect to the subtraction operators presented in \cite{Armadillo:2025mfx} for the general EW case, we restricted the sum of Equation~(\ref{eq:I02}) only to the massive fermions considered in the computation, with multiplicity $n_k$ (3 for the coloured quarks, 1 for the leptons), electric charge $q_k$, and mass $m_k$. In addition, while the definition of $\Gamma'$ implies $\Gamma'_1$ to represent a purely fermionic contribution to the subtraction operator, the specification of $\Gamma_{1,\text{fer}}$ is necessary to exclude the purely bosonic contributions that appear in the full EW expression of $\Gamma_1 = \Gamma_{1,\text{bos}} + \Gamma_{1,\text{fer}}$. We report here its definition:
\begin{align}
    \Gamma_{1,\text{fer}} = \sum_{k}^{\text{massless}} n_k q_k^2 \ \Bigg[&\sum_i \left(\frac{130}{27}+\frac{2\pi^2}3\right) Q_i^2+ \sum_j  \frac{40}{9} Q_j^2 -\frac{80}{9} \Bigg( Q_1 Q_2 \log\left(\frac{\mu_R^2}{- 2p_{1}\cdot p_{2} }\right) \nonumber\\
    &-\sum_{j_1 \neq j_2} \frac{Q_{j_1} Q_{j_2}}2 \beta_{j_1, j_2}\coth\beta_{j_1, j_2}- \sum_{i,j} Q_i Q_j\log\left(\frac{\mu_R\, m_j}{ 2 p_i\cdot p_j}\right)\Bigg) \Bigg] \;.
\end{align}
We refer to massless initial state particles labeled with $i=1,2$, with momentum $p_i$ and electric charge $Q_i$, while with the labels $j=3, \dots, n+2$ we refer to the massive final state emitters, with mass $m_j$, momentum $p_j$ and electric charge $Q_j$. We label with $\beta_{j_1, j_2}$ the function defined by the relation $\cosh \beta_{j_1, j_2}= -\left(p_{j_1}\cdot p_{j_2}\right)/\left(m_{j_1} m_{j_2}\right)$. The physically allowed values for $j_1$ and $j_2$ outgoing are $\cosh \beta_{j_1, j_2} \leq 1$, corresponding to $\beta_{j_1, j_2}=-b +i\pi$ with real $b\geq0$.
The expression of ${\cal I}^{(0,2)}_{\rm fer}$ receives a contribution from the renormalisation of the one-photon emission vertex, which has been performed in the $\overline{MS}$ scheme, and contributes with a $\beta_0$ term, which reads \begin{align}
     \label{eq:beta0QED}
     \beta_0=&-\frac{4}{3} \, \sum_{k}^{\text{massless}} n_k \, q_k^2 \;.
\end{align}

\section{Prescription for \texorpdfstring{$\gamma_5$}{gamma5}}
\label{sec:gamma5}

The presence of chiral couplings in the EW theory is a well known problem for calculations performed in dimensional regularisation, since they require an extension of the inherently 4-dimensional $\gamma_5$ operator to the $d$-dimensional case. It can be shown that such an extension can not simultaneously preserve the three following properties: the anti-commutativity with the Dirac $\gamma^\mu$ matrices
\begin{equation}
 \{ \gamma_{\mu}, \gamma_5 \} = 0   \ ,
 \label{eq:anticommutativity}
\end{equation}
the trace condition 
\begin{equation}
    \text{Tr}(\gamma_\mu \gamma_\nu \gamma_\rho \gamma_\sigma \gamma_5) = 4i\epsilon_{\mu\nu\rho\sigma} \, ,
    \label{eq:tracecondition}
\end{equation}
and the invariance of the trace of the product of Dirac matrices and $\gamma_5$ under cyclic permutations.
As a consequence, two main strategies have been historically employed to perform EW computations featuring $\gamma_5$ matrices in dimensional regularisation: the HVBM's $\gamma_5$ scheme (\cite{'tHooft:1972fi,Breitenlohner:1975hg,Breitenlohner:1976te,Breitenlohner:1977hr} and the Kreimer's scheme (\cite{Kreimer:1989ke,Korner:1991sx,Kreimer:1993bh}), respectively preserving the trace cyclicity or the anti-commuting relation.

In our framework, we use the Kreimer scheme as implemented in the Mathematica package {\sc Abiss} \cite{ABISS}.
{\sc Abiss} contains an automatic implementation of the strategy that has been presented in \cite{Heller:2020owb} for the computation of the Drell-Yan mixed QCD-EW corrections: to perform the required Dirac algebra,
it preserves the anti-commutativity Equation (\ref{eq:anticommutativity});
it adopts a suitable definition of the trace operator consistent with Equation (\ref{eq:tracecondition}), while it abandons the cyclicity of the trace, introducing a reading point for each distinct trace present in the amplitude. 
This choice significantly reduces the computational load of the calculation: the anti-commuting property, together with the relation $\gamma_5^2=1$, allows to write each Dirac trace with at most a single $\gamma_5$ matrix in its rightmost position.
The prescription for the definition of such traces employs a projection of the Lorentz quantities onto 4- and $(-2\epsilon)$-dimensional subspaces: specifically, the Levi-Civita tensor $\varepsilon_{\mu_1 \mu_2 \mu_3 \mu_4}$ and $\gamma_5$ represent  4-dimensional objects, satisfying the relations:
\begin{align}
    &\epsilon_{\mu\nu\rho\sigma} \epsilon^{\mu\nu\rho\sigma} = -24 \ ,\\ \label{eq:gamma5sub}
    & \gamma_5=\frac 1{4!} \varepsilon_{\mu_1 \mu_2 \mu_3 \mu_4}\gamma^{\mu_1}\gamma^{\mu_2}\gamma^{\mu_3}\gamma^{\mu_4}\; \ .
\end{align}
The product of two Levi-Civita tensors is expressed as a combination of 4-dimensional metric tensors \cite{Heller:2020owb}.
The integrals with a scalar product involving a 4-dimensional metric tensor are replaced by a combination of integrals with  $d$- and ($-2\epsilon$)-dimensional scalar products in the respective numerators.
The latter are in turn expressed as linear combinations of new scalar integrals of the same integral families
evaluated in an increased number of space-time dimensions, up to $d=6-2\epsilon$ and $d=8-2\epsilon$ for a two-loop computation. 

Each interference term, before the UV renormalisation and the subtraction of the IR divergences, depends on the details of the above prescription, namely the choice of both the ordering of the spinor chains within a fermionic trace and the reading point for the internal closed fermionic loops.
We have explicitly checked that the UV-renormalised and IR-subtracted interference terms, on the contrary, do not depend on the adopted convention described above.

\section{Production of a muon pair in quark-antiquark annihilation}
\label{sec:amplitude}

\subsection{The scattering amplitude and the fermionic corrections}
We consider the production of a  pair of massive leptons in quark-antiquark annihilation, more specifically
an up-quark pair which annihilates into an muon pair 
\begin{equation}
 u(p_1) + \bar{u}(p_2) \rightarrow \mu^+(p_3) + \mu^{-} (p_4) \,.
\label{eq:process}
\end{equation} 
The Mandelstam variables are defined as:
\begin{equation}
 s = (p_1+p_2)^2, \,\  t = (p_1-p_3)^2, \, \ u = (p_2-p_3)^2 \,\,\, {\rm with} \,\, s+t+u= 2 m_\mu^2 \,,
\end{equation}
with $m_\mu$ the muon mass.
The on-shell conditions of the external particles are:
\begin{equation}
 p_1^2 = p_2^2 =  0; ~ ~ ~ p_3^2 = p_4^2  = m_\mu^2.
\end{equation}
We write the virtual amplitude for the production of a pair of massive muons as
\begin{equation}
\label{eq:expansion}
    |\mathcal{M}_0 (u\bar u\to \mu^+\mu^-)\rangle = 4\pi\alpha_0 \sum_{j=0}^{\infty} \left( \frac{\alpha_0}{4\pi}\right)^j |\mathcal{M}_0^{(0,j)}(s,t,m_{\mu,0},m_{f,0},\mu_{W,0},\mu_{Z,0},\mu_{H,0})\rangle\;,
\end{equation}
where $|{\cal M}_0^{(0,j)}\rangle$ is the bare amplitude at $j^{th}$ perturbative order in $\alpha_0$, and all the parameters (the electric charge $e_0$, the muon mass $m_{\mu,0}$, the mass $m_{f,0}$ of a generic fermion $f$, the masses of $W$,$Z$,$H$ bosons $\mu_{(W,Z,H),0}$) and fields are treated as bare quantities, labeled with a index 0. 
The renormalised amplitude is obtained by the replacement of all the bare quantities with their renormalised counterparts, 
as described in detail in Section \ref{sec:renormalization}:
\begin{align}
&|{\cal M}(u\bar u\to \mu^+\mu^-)\rangle~=~
Z^{\frac12}_u Z^{\frac12}_{\bar u} Z^{\frac12}_{\mu^+} Z^{\frac12}_{\mu^-}~16\pi^2~\times\nonumber\\
&~~~~~~~~~~~~~~~~~~~~~~~~~~~~~\sum_{j=0}^\infty \left(\frac{e (1+\delta Z_e)}{4\pi}\right)^{2j+2} |\hat{\mathcal{M}}^{(0,j)}(s,t,m_{\mu,f}+\delta m_{\mu,f}, \mu^2_{W,Z,H}+\delta \mu^2_{W,Z,H})\rangle\nonumber\\
&~~~~~~~~~~~~~~~~~~~~~~~~~= \ 4\pi\alpha\sum_{j=0}^\infty \left(\frac{\alpha}{4\pi}\right)^{j} |\mathcal{M}^{(0,j)}(s,t,m_{\mu},m_f,\mu^2_W,\mu^2_Z,\mu^2_H)\rangle\;.
\label{eq:amprenormalised}
\end{align}
Since not only the bare amplitude, but also the UV counterterms are computed in perturbation theory as power series in the bare coupling $e_0$, we expand the renormalised amplitude and  collect all the terms of the same order in the coupling constant, checking the cancellation of the UV divergences.
The result can be eventually cast as an expansion in the renormalised coupling $\alpha=e^2/(4\pi)$, where $|\mathcal{M}^{(0,j)}\rangle$ includes diagrammatic and counterterm contributions and is UV finite.
We write the UV-renormalised amplitude as a Laurent expansion in $\epsilon$, and we keep only terms up to 
$\mathcal{O}(\epsilon^2)$ or $\mathcal{O}(\epsilon^0)$, at one- or two-loop respectively, in order to include all the second-order finite contributions when we perform the IR subtraction described in Section~\ref{sec:infra}.
\begin{figure}
\begin{minipage}{0.24\textwidth}
\centering
\includegraphics[width=0.85\textwidth]{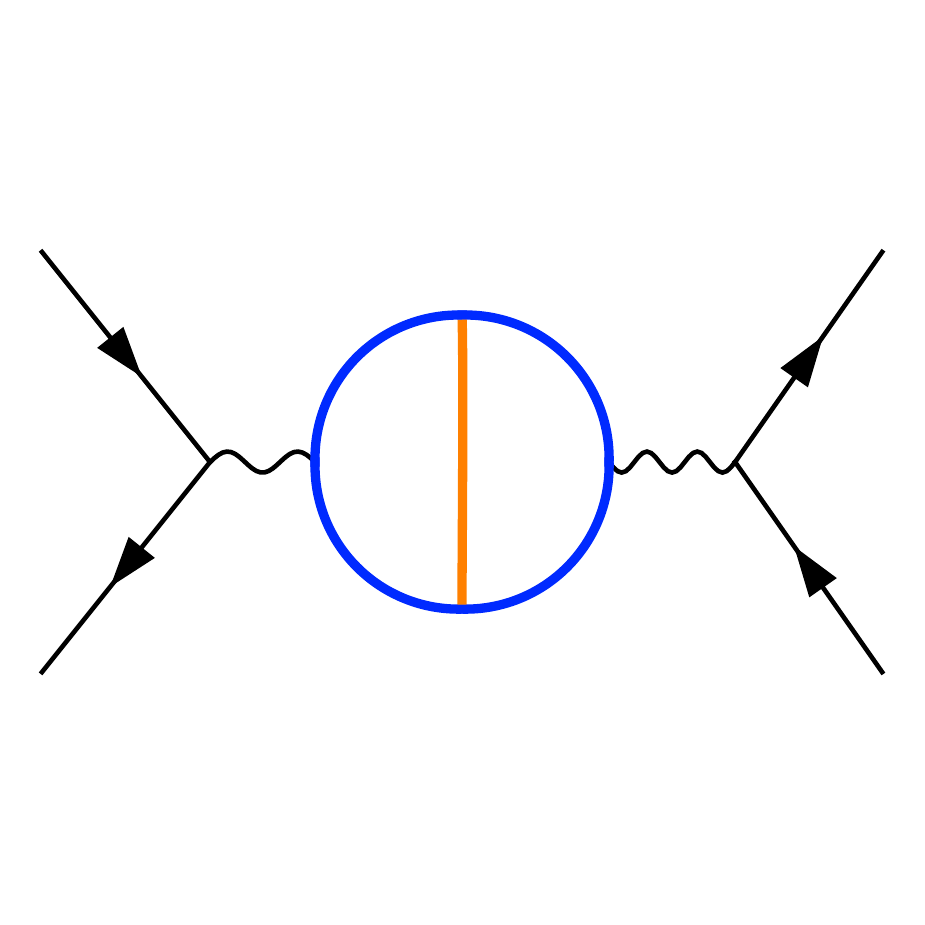} \subcaption{}
\end{minipage}
\begin{minipage}{0.24\textwidth}
\centering
\includegraphics[width=0.85\textwidth]{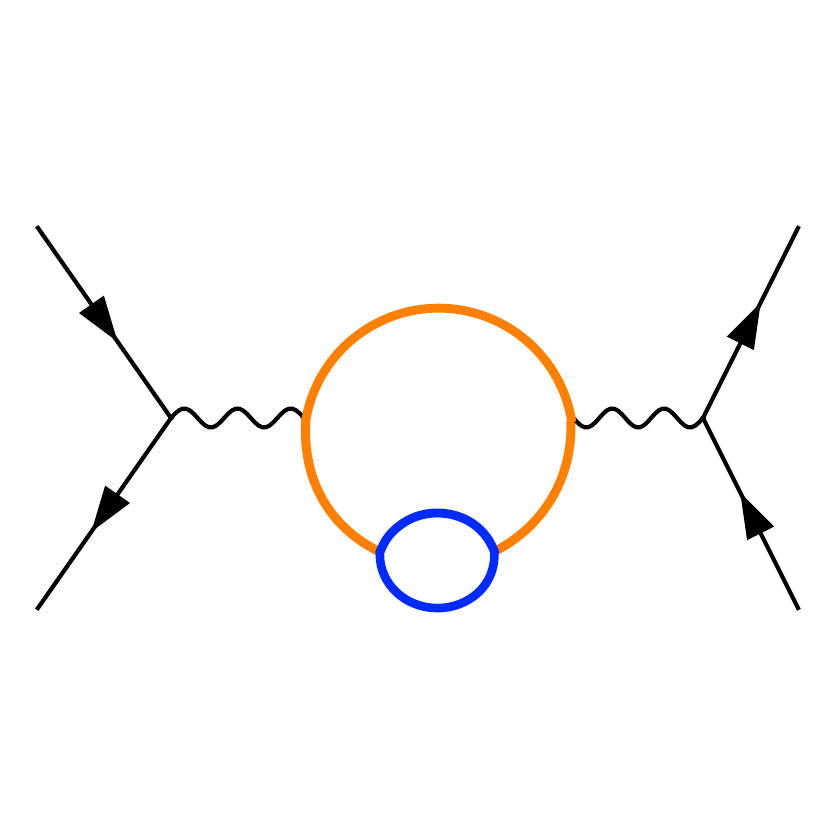}
\subcaption{}\end{minipage}
\begin{minipage}{0.24\textwidth}
\centering
\includegraphics[width=0.85\textwidth]{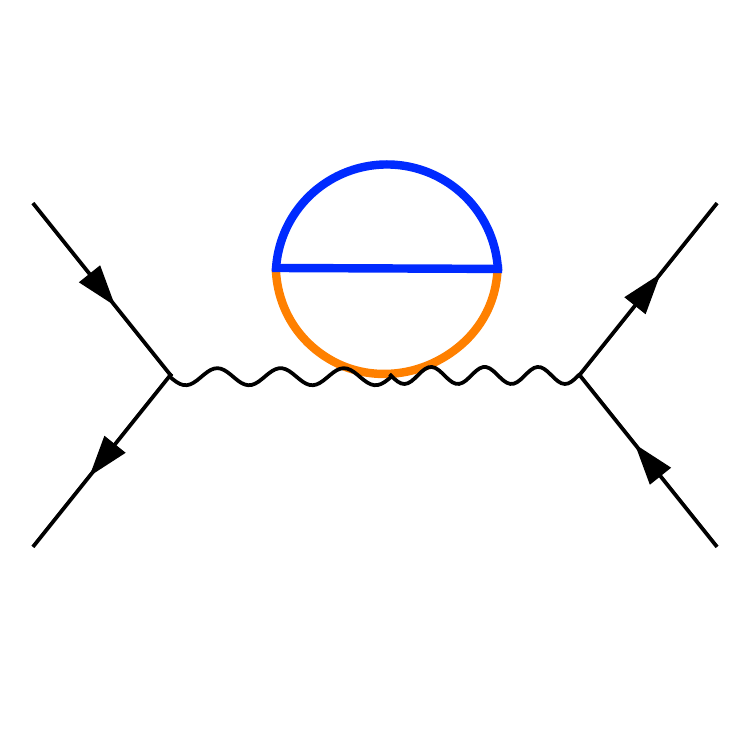}
\subcaption{}
\end{minipage}
\begin{minipage}{0.24\textwidth}
\centering
\includegraphics[width=0.85\textwidth]{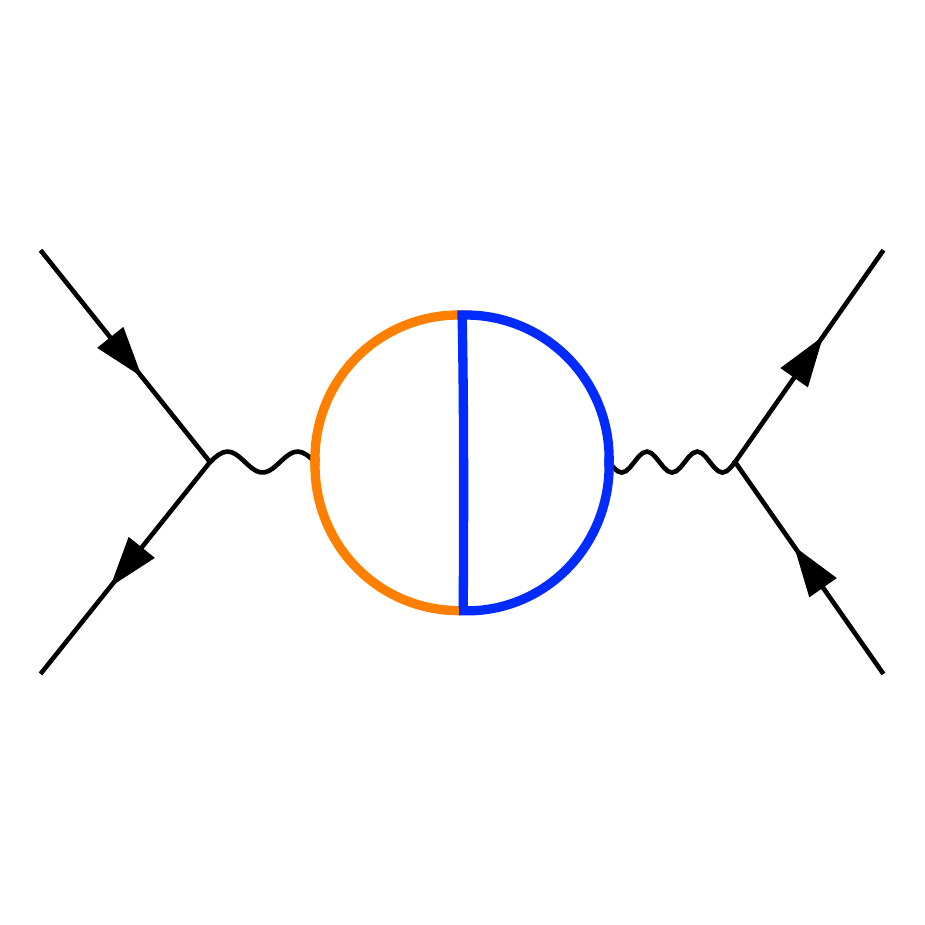}
\subcaption{}
\end{minipage}
\begin{minipage}{0.24\textwidth}
\centering
\includegraphics[width=0.85\textwidth]{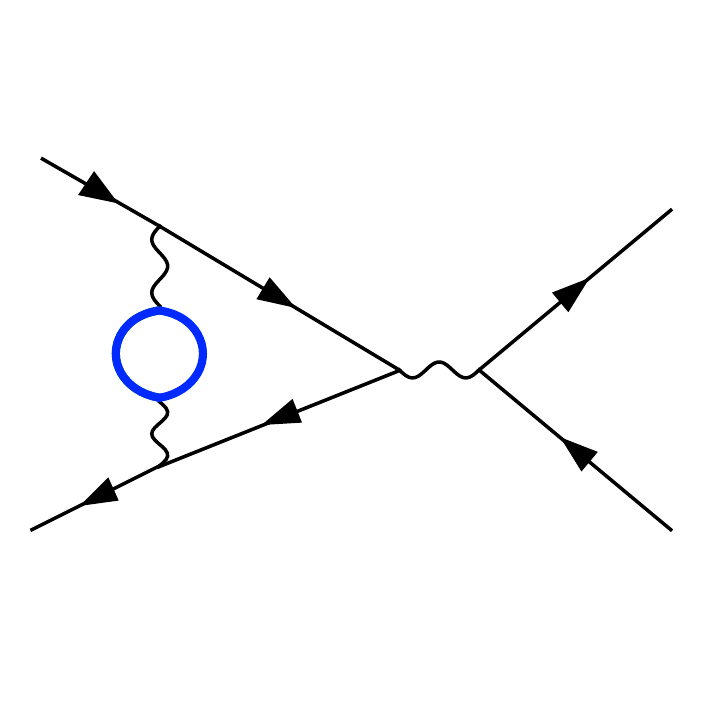}
\subcaption{}
\end{minipage}
\begin{minipage}{0.24\textwidth}
\centering
\includegraphics[width=0.85\textwidth]{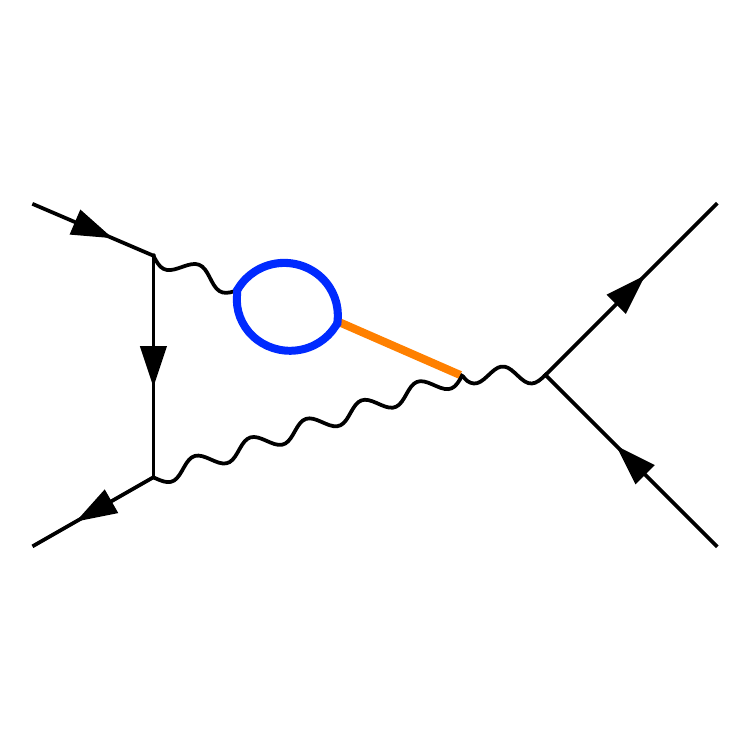}
\subcaption{}
\end{minipage}
\begin{minipage}{0.24\textwidth}
\centering
\includegraphics[width=0.85\textwidth]{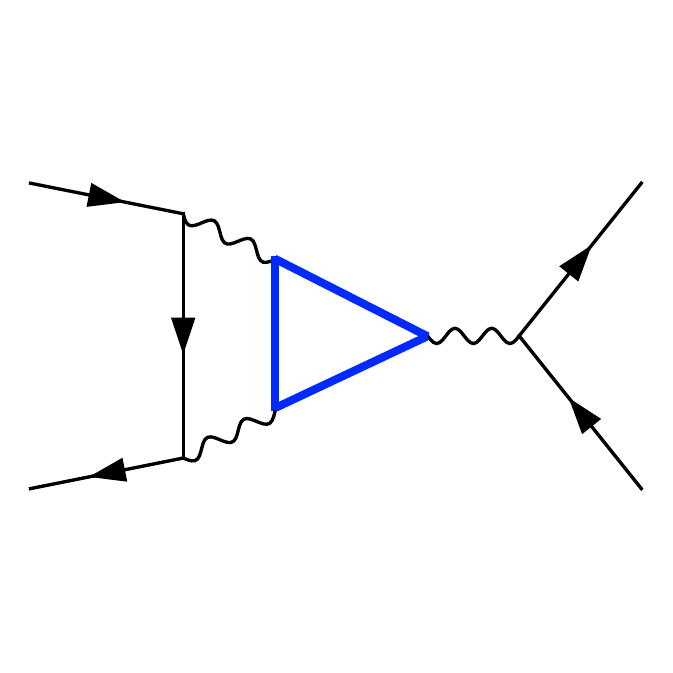}
\subcaption{}
\end{minipage}
\begin{minipage}{0.24\textwidth}
\centering
\includegraphics[width=0.85\textwidth]{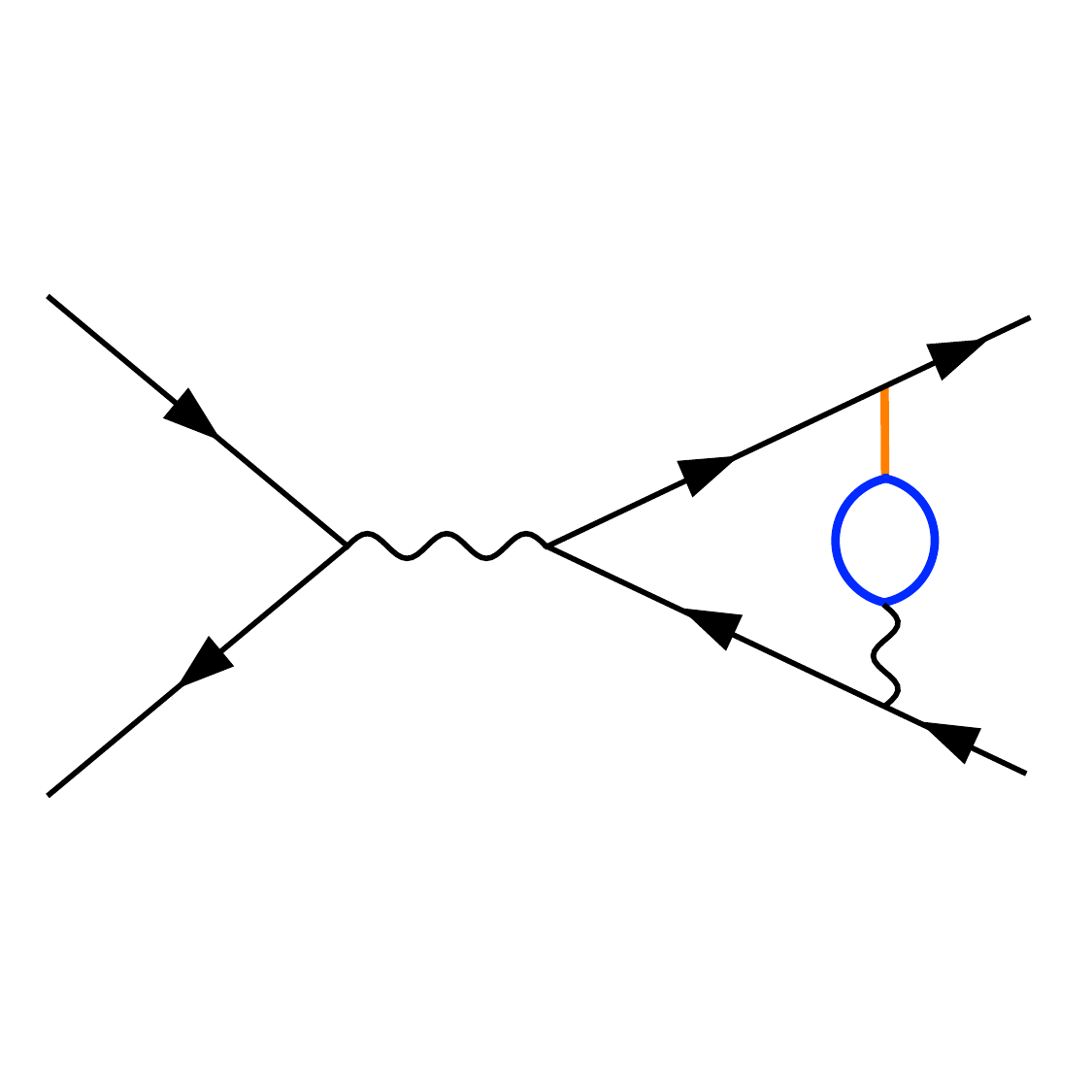}
\subcaption{}
\end{minipage}
\begin{minipage}{0.24\textwidth}
\centering
\includegraphics[width=0.85\textwidth]{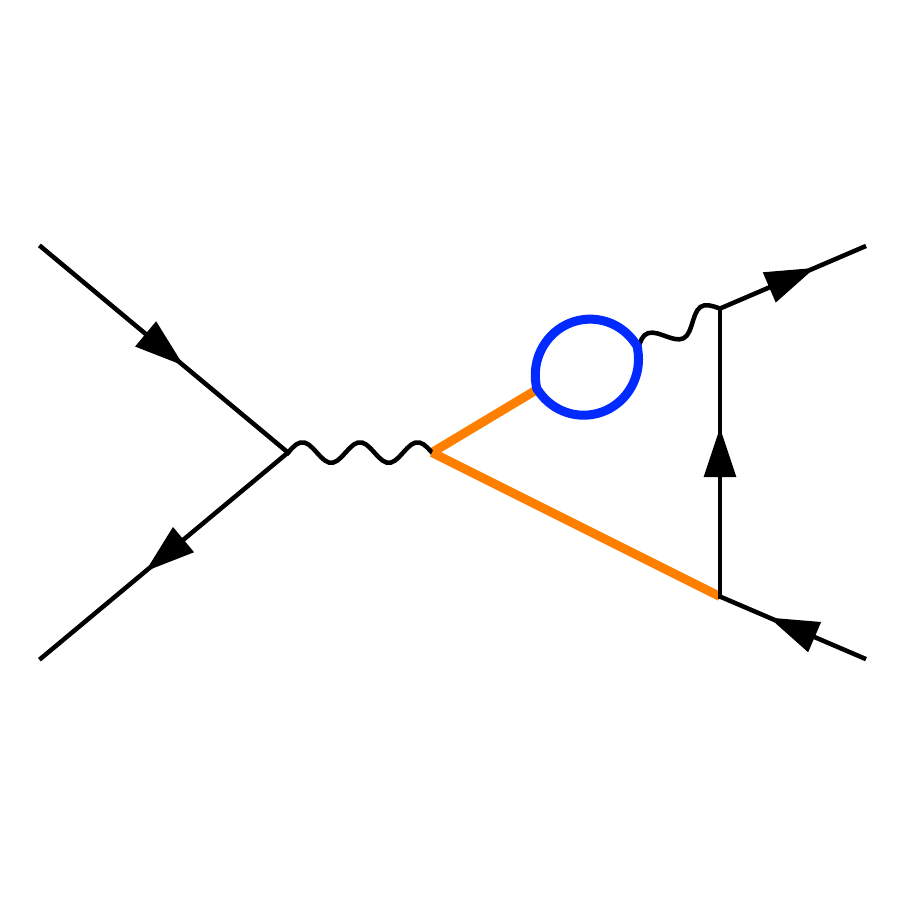}
\subcaption{}
\end{minipage}
\begin{minipage}{0.24\textwidth}
\centering
\includegraphics[width=0.85\textwidth]{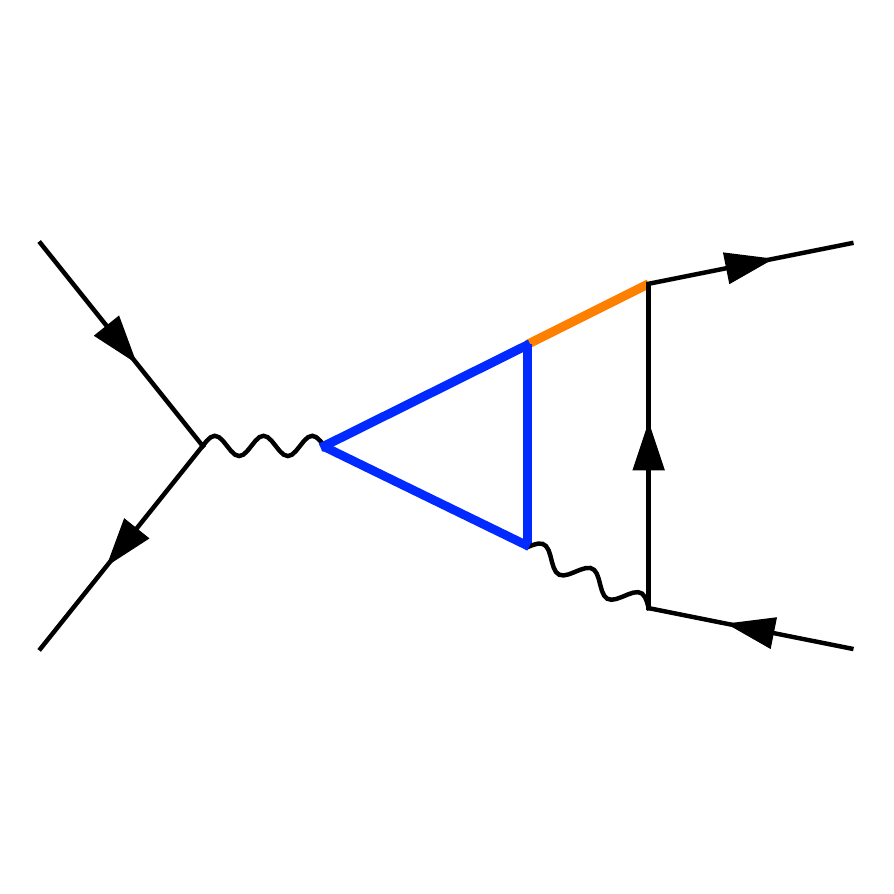}
\subcaption{}
\end{minipage}
\begin{minipage}{0.24\textwidth}
\centering
\includegraphics[width=0.85\textwidth]{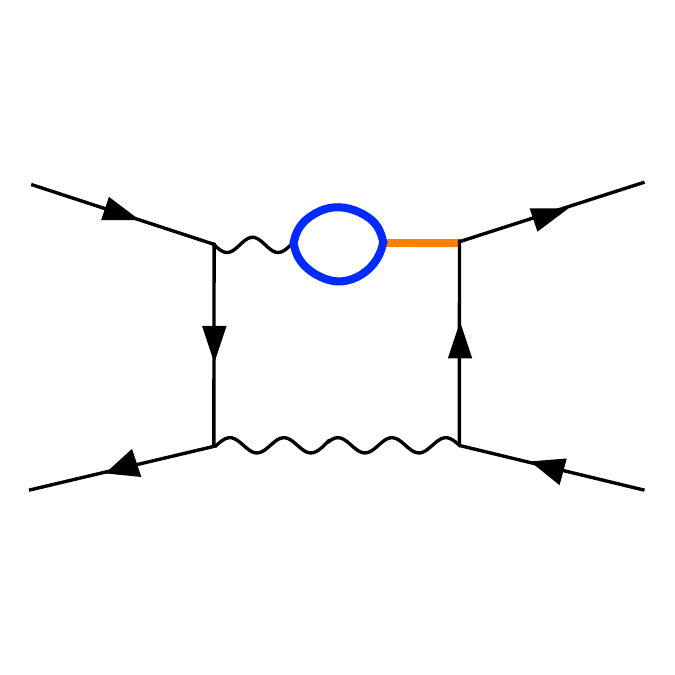}
\subcaption{}
\end{minipage}
\begin{minipage}{0.24\textwidth}
\centering
\includegraphics[width=0.85\textwidth]{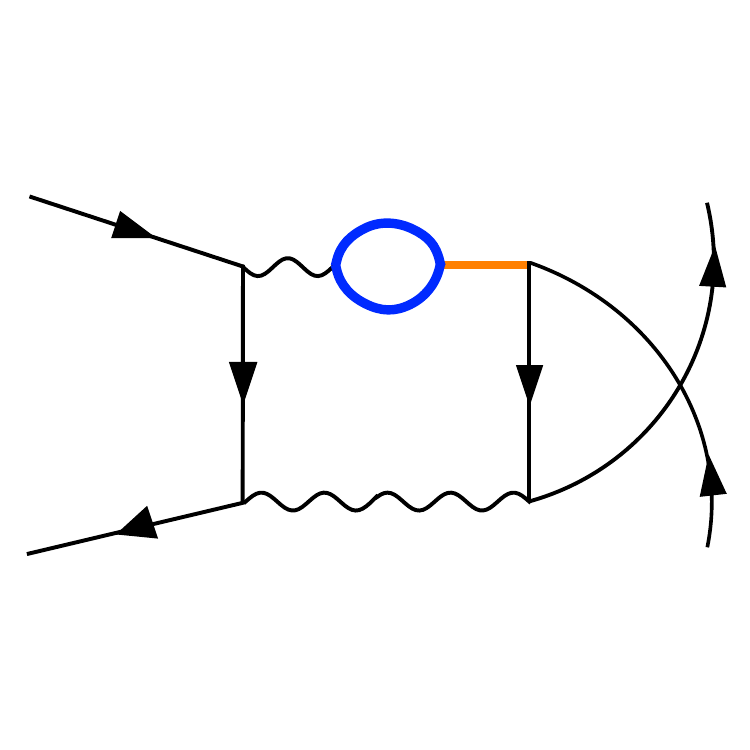}
\subcaption{}
\end{minipage}
\begin{minipage}{0.24\textwidth}
\centering
\includegraphics[width=0.85\textwidth]{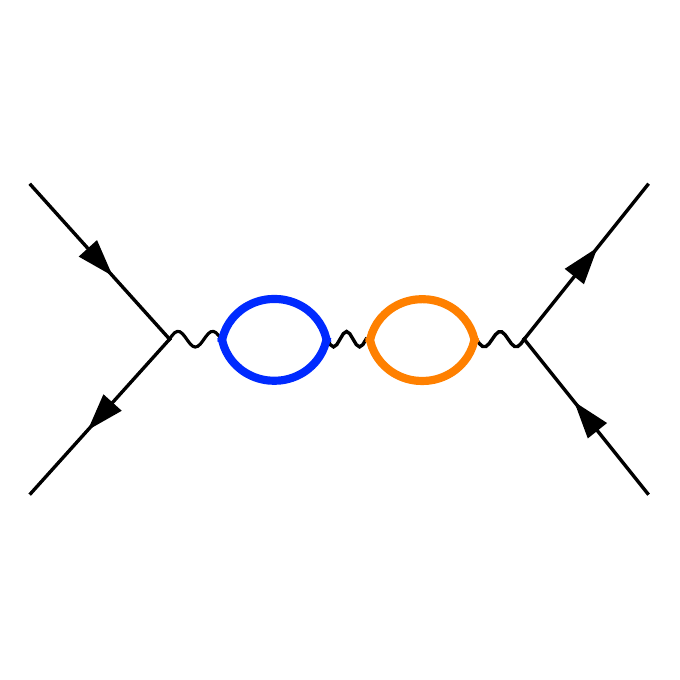}
\subcaption{}
\end{minipage}
\begin{minipage}{0.24\textwidth}
\centering
\includegraphics[width=0.85\textwidth]{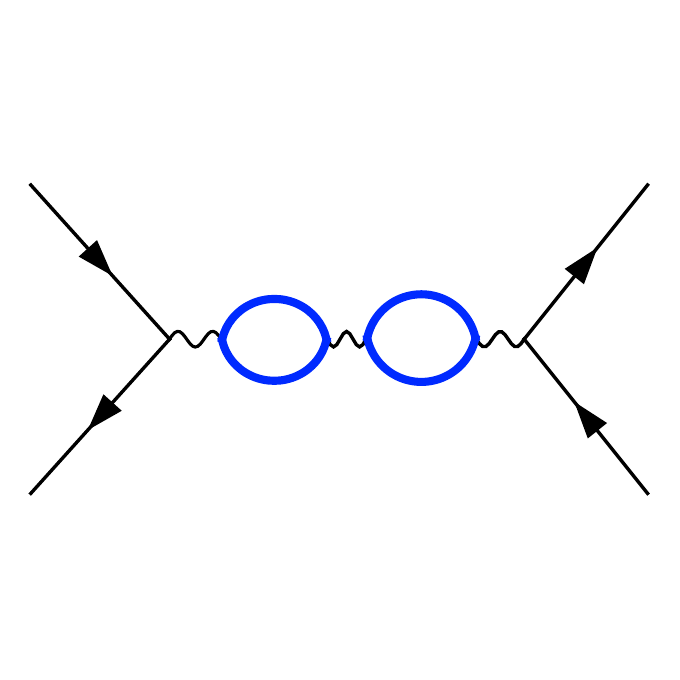}
\subcaption{}
\end{minipage}
\begin{minipage}{0.24\textwidth}
\centering
\includegraphics[width=0.85\textwidth]{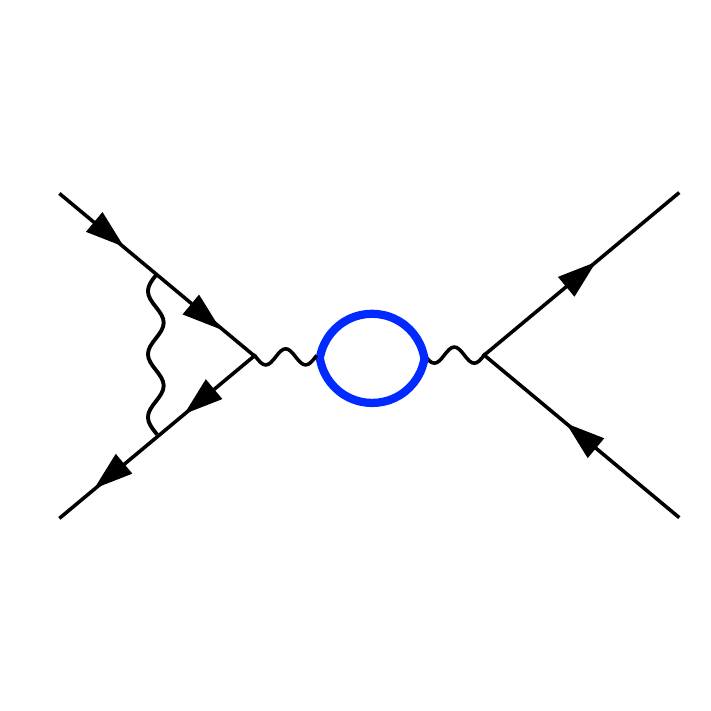}
\subcaption{}
\end{minipage}
\begin{minipage}{0.24\textwidth}
\centering
\includegraphics[width=0.85\textwidth]{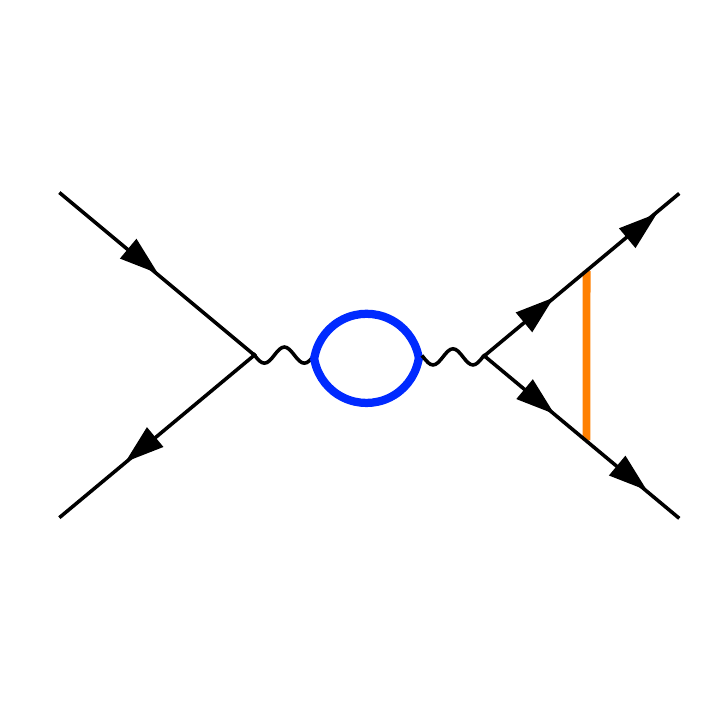}
\subcaption{}
\end{minipage}
\caption{\label{fig:sample2Ldiagrams} {Sample of the two-loop bare fermionic diagrams: one-particle irreducible and factorisable one-loop$\times$one-loop contributions. The blue lines represent internal fermions, the orange lines generic internal bosons (vectors or scalars). The wavy lines represent vector boson propagators.  }}
\end{figure}
\begin{figure}
\begin{minipage}{0.245\textwidth}
\centering
\includegraphics[width=0.85\textwidth]{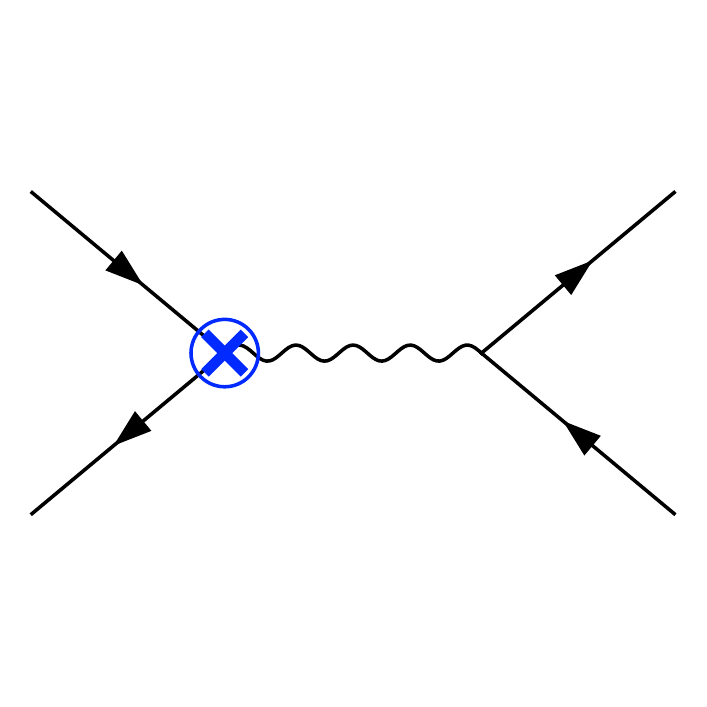}\subcaption{}
\end{minipage}
\begin{minipage}{0.245\textwidth}
\centering
\includegraphics[width=0.85\textwidth]{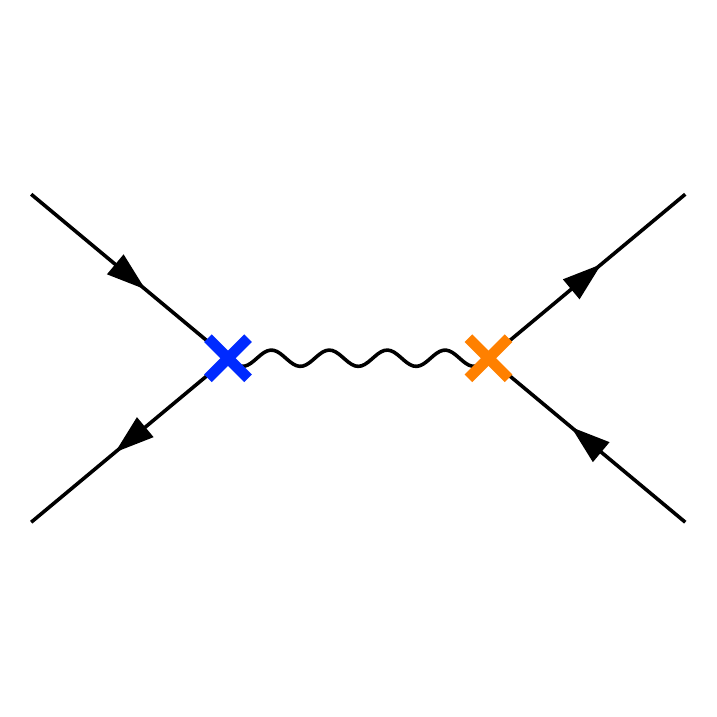}\subcaption{}
\end{minipage}
\begin{minipage}{0.245\textwidth}
\centering
\includegraphics[width=0.85\textwidth]{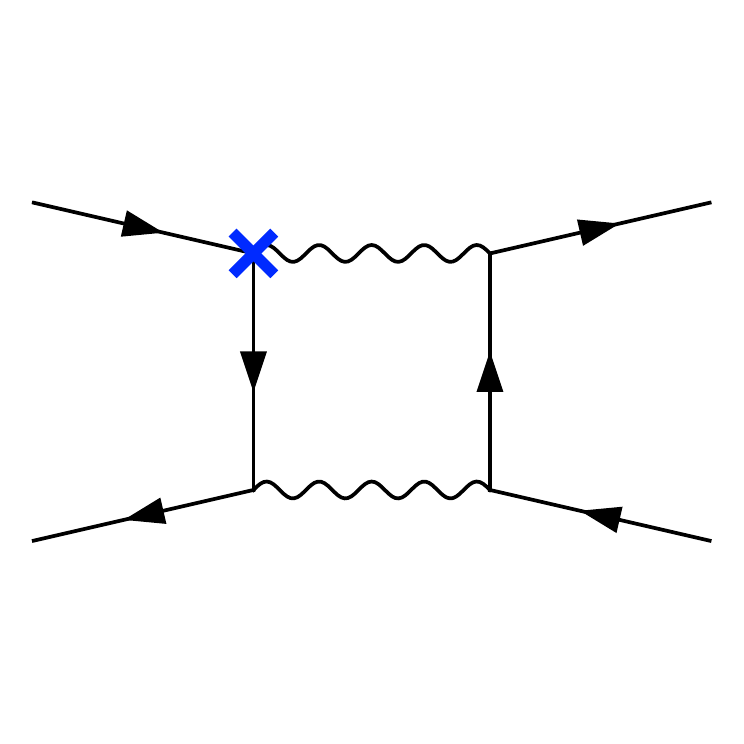}\subcaption{}
\end{minipage}
\begin{minipage}{0.245\textwidth}
\centering
\includegraphics[width=0.85\textwidth]{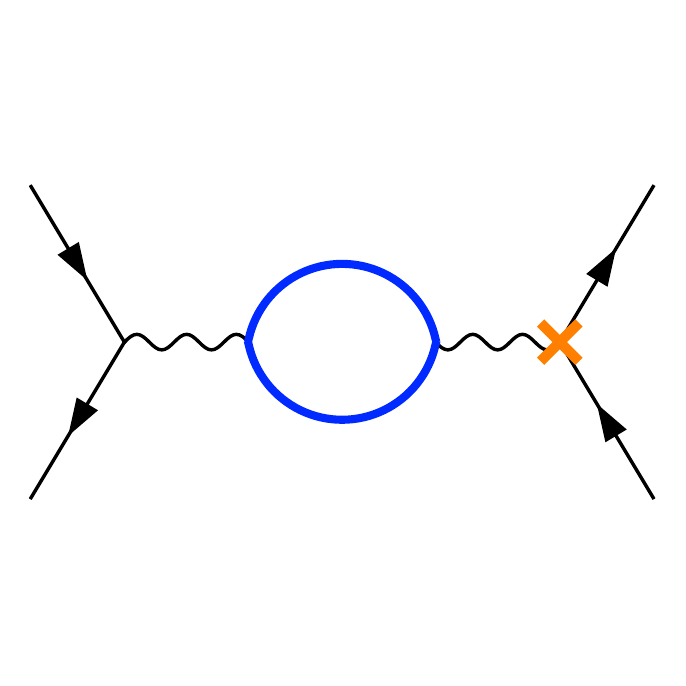}\subcaption{}
\end{minipage}
\caption{\label{fig:fermionloopandCTs} {Sample of the Feynman diagrams with the insertion of first (cross) and second (cross and circle) order counterterms, we indicate in blue the fermionic ones and in orange the bosonic ones. The thick blue lines represent internal fermions, the wavy lines represent vector bosons.}}
\end{figure}

We regularise all the soft IR divergences and the initial state collinear divergences in dimensional regularisation. The final-state collinear singularities are regularised by the muon mass.  Our calculation retains the exact logarithmic accuracy with respect to the logarithms $\log(m_\mu^2/s)$.
In order to reduce the computational load, we approximate the amplitude by treating the muon as massless in all those diagrams which can not feature a final-state IR divergence. Because of this choice, we expect our final results to be exact up to power corrections in the muon mass.

In the following, we focus on the  second order fermionic EW corrections to the process $u\bar u\to\mu^+\mu^-$, namely all those terms characterised by the presence of at least one closed fermionic loop in the Feynman diagrams or in the counterterms contributing to the scattering amplitude. The fermionic contributions form a gauge-invariant subset of the virtual corrections to the two-loop amplitude.  We always sum over all the members of a fermionic family, leptons and quarks, leading to the cancellation of all the anomalous contributions, and we include the three known fermionic families.
The evaluation of the fermionic corrections to the two-loop scattering amplitude of a process implies that in the definition of the counterterms 
we distinguish fermionic and bosonic contributions.
Calling $p_0$ one of the bare parameters of the SM lagrangian, we replace it with its renormalised counterpart and an appropriate counterterm $p_0\to p+\delta p$. The latter admits a perturbative expansion in the coupling constant $\alpha$ 
\begin{equation}
    \delta p = \sum_{i=1}^\infty \alpha^i \delta p^{(i)}
    = \sum_{i=1}^\infty \alpha^i \left(\delta p_{\text{\emph{fer}}}^{(i)}+\delta p_{\text{\emph{bos}}}^{(i)}\right)
\end{equation}
where we have split, at each perturbative order, the counterterm into a fermionic and a purely bosonic contributions, labelled by the index \emph{fer} or \emph{bos}; the former contains at least one closed fermionic loop while the latter is free from such factors.

In Figure \ref{fig:sample2Ldiagrams} we show the relevant topologies  of the two-loop Feynman diagrams contributing to the bare amplitude term $ | \unM^{(0,2)}_{\rm fer} \rangle$.
We find four distinct cases of two-loop one-particle irreducible (1PI) self-energy corrections, with the fermion loop attached to the external gauge boson, \ref{fig:sample2Ldiagrams}-(a), inserted along an internal boson line, \ref{fig:sample2Ldiagrams}-(b,c), or forming a triangle attached to one external gauge boson, \ref{fig:sample2Ldiagrams}-(d).
We find four distinct cases of two-loop 1PI initial-state vertex corrections, with the fermion loop  inserted along an internal boson line, \ref{fig:sample2Ldiagrams}-(e,f), or forming a triangle attached to the external gauge boson, \ref{fig:sample2Ldiagrams}-(g).
In the final state case we have analogous topologies \ref{fig:sample2Ldiagrams}-(h,i,j). The massive nature of the final-state muon pair allows additional contributions with the exchange of an Higgs or a pseudoscalar boson, or also a (pseudo)scalar-vector mixing. 
We find two-loop 1PI box corrections, with the fermion loop inserted along an internal boson line, and the underlying one-loop box having a planar \ref{fig:sample2Ldiagrams}-(k) or a crossed \ref{fig:sample2Ldiagrams}-(l) routing of the momenta.
Finally, we consider also the product of two one-loop factors such as the consecutive insertion of two one-loop self-energies (at least one fermionic) in the tree-level diagram, \ref{fig:sample2Ldiagrams}-(m,n), or the product of a one-loop vertex corrections with a fermionic self-energy, \ref{fig:sample2Ldiagrams}-(o,p).
In Figure \ref{fig:fermionloopandCTs} we show some representative examples of diagrams with counterterm insertions. 
We consider the insertion of a two-loop fermionic counterterm on a tree-level diagram \ref{fig:fermionloopandCTs}-(a), 
the insertion of two one-loop counterterms on a tree-level diagram, with at least one being of fermionic kind, \ref{fig:fermionloopandCTs}-(b), 
the insertion of a one-loop fermionic counterterm on a one-loop bosonic diagram \ref{fig:fermionloopandCTs}-(c) and
the insertion of a one-loop bosonic counterterm on a one-loop fermionic diagram \ref{fig:fermionloopandCTs}-(d).
The general expression in Equation (\ref{eq:amprenormalised}) for the UV-renormalised amplitude can be specialised to the fermionic case, by requesting the presence in each term of the sum of at least one closed fermionic loop, be it in the Feynman diagrams or in the expression of the counterterms.


\subsection{Master Integrals}

In this Section, we report on the reduction of the two-loop Feynman integrals to Master Integrals and their subsequent numerical evaluation. Both steps were fully automated within the \oceann~framework presented in \cite{Armadillo:2025mfx}. Specifically, the reduction was performed using \kira~\cite{Lange:2025fba} in combination with {\sc FireFly} \cite{Klappert:2020aqs}, while the numerical evaluation was carried out with \amflow~\cite{Liu:2022chg} and \seasyde~\cite{Armadillo:2022ugh}.

To simplify the calculation, we neglected the mass of the final-state muon whenever possible. In particular, the Master Integrals exhibit both soft and collinear IR divergences. While soft divergences are handled via dimensional regularisation, final-state collinear singularities are regularised by retaining a finite muon mass. However, keeping $m_\mu\neq 0$ introduces an additional kinematic scale, significantly slowing down the reduction procedure.
To mitigate this computational bottleneck, we exploit the fact that final-state collinear divergences only arise when a photon couples to a final-state muon. For instance, the diagrams in Figure~\ref{fig:sample2Ldiagrams}-(h-l) develop an IR collinear singularity only if the exchanged boson is a photon. If, instead, we replace it by a massive gauge boson, the singularity is absent. Consequently, in the latter case, we can safely set $m_\mu=0$.
The error introduced by neglecting the mass of the muon is suppressed by $\mathcal{O}(m_\mu^2/\mu_W^2)$ factor, which is negligible for any phenomenological applications. This strategy effectively bounds the number of mass scales, lowering the computational load of the calculation.

The specific topologies we encounter include two-loop self-energies with up to five massive lines and two distinct masses, arising from the diagrams shown in Figure~\ref{fig:sample2Ldiagrams}-(a-d). For the two-loop triangles, such as those in Figure~\ref{fig:sample2Ldiagrams}-(e-j), we compute integrals with up to five massive lines and four different masses. 
Finally, we evaluate topologies consisting of one-loop boxes with a one-loop bubble insertion on an internal line, as depicted in Figure~\ref{fig:sample2Ldiagrams}-(k,l). These feature at most six massive propagators and three different masses.

Compared to the calculation of the two-loop mixed QCD-EW corrections presented in \cite{Armadillo:2022bgm,Armadillo:2024nwk}, the integrals considered here involve a higher number of internal mass scales, going from two to three for the boxes, and up to four for the triangles. Conversely, the requirement of a closed fermion loop restricts the maximum number of propagators in our two-loop boxes to six, whereas in \cite{Armadillo:2022bgm,Armadillo:2024nwk} there were seven. Consequently, the overall complexity of the calculation, from the perspective of integral reduction, remains comparable.

The most demanding topology, i.e. a two-loop box with six massive propagators and three distinct masses, required approximately 4 hours for the reduction on an 96-core machine with 500 GB of memory. The subsequent numerical evaluation in one point of the phase-space with \amflow~ took roughly 1 hour and 45 minutes on 96 cores.

\section{Results}
\label{sec:results}
We present in this Section the outcome of our calculation, based on the results illustrated in Sections  \ref{sec:renormalization}, \ref{sec:infra}, and \ref{sec:gamma5}.
The automation of the evaluation of the individual interference terms between two-loop virtual and tree-level diagrams  mostly relies on the recent developments of the package \abiss\, \cite{ABISS}.
The evaluation of the Feynman Master Integrals is based on the packages \amflow~\cite{Liu:2022chg} and \seasyde~\cite{Armadillo:2022ugh}.
This complete workflow is executed within the framework \oceann~\cite{Armadillo:2025mfx}; the resulting automation of the calculation allows to perform systematic tests and  validation  of the results.

Among the technical checks, the cancellation of the poles in $\epsilon$, which represent in dimensional regularisation the UV and IR singularities, is a basic necessary condition. At second order we have the entanglement of UV and IR poles, such that the two groups of divergences can not be discussed in separate steps. In addition, the prescription for the $\gamma_5$ treatment generates, in general, spurious poles (and also spurious finite contributions) which cancel at the level of the UV-renormalised IR-subtracted amplitude. The necessary check of the poles cancellation can thus successfully take place only when the three ingredients mentioned above are all correctly implemented.

The second-order EW corrections feature  terms developing an IR divergence which does not factorise from the rest of the Feynman diagram expression. If the associated Feynman integral can not be solved in closed form in terms of elementary functions, it is often difficult or impossible to express the IR-divergent component as an analytical term.
These cases have motivated us \cite{Armadillo:2025mfx} to pursue the numerical cancellation of all the poles, relying on an arbitrary precision approach (i.e. with finite precision which can be set at an arbitrarily large value by the user). The algorithm implemented in \seasyde\, delivers such high-precision numbers; in this language the cancellation is verified if we obtain a coefficient equal to $10^{-P+L}$ with an arbitrary positive integer value of $P$ and taking into account a precision loss $L$. 
The same numerical approach is used to verify the Ward identities satisfied by the Green's function appearing in the amplitude.

We remark that in the calculation we have approximated the amplitude by keeping the muon mass in the two-loop Feynman diagrams which feature a final state collinear divergence (a logarithm of the muon mass), and set it to zero otherwise. This choice leads to incomplete expressions of the coefficients of the $\epsilon$ poles, because of missing terms proportional to $m_\mu^2/\mu_W^2$. As a consequence we observe a cancellation of the $\epsilon$ poles only at the $10^{-7}$ level, i.e. up to terms proportional to the squared muon mass.
We have checked that these are all and the only terms responsible for the imperfect cancellation, by reducing the numerical value of the muon mass by a factor 100, observing in turn the cancellation improving by a factor $100^2$, consistent with the expected squared power-like behavior. 

\subsection{Numerical results}
\label{sec:numerical}
In Table \ref{tab:ircanc} we present the value of the coefficient of the Laurent expansion of the UV-renormalised and IR-subtracted amplitude, at one given kinematic point.
For each power of $\epsilon$ we have three rows corresponding to the diagrammatic and the two IR subtraction contributions of Equation (\ref{eq:subtracted}). For each negative power, we can check that summing the values in the three rows gives zero with a relative precision of $10^{-7}$, as discussed in the previous Section.
The results are obtained using the following values for the masses and Mandelstam invariants:
\begin{center}
 \begin{tabular}{l l l} 
 \hline\hline
  $s=10000~\mathrm{GeV}^2$ &\hspace{20pt} $m_e$ = 0.510998 MeV &\hspace{20pt} $m_Z$ = 91.1535 GeV \\
  $t=-2500~\mathrm{GeV}^2$ &\hspace{20pt} $m_\mu$ = 0.105660 GeV &\hspace{20pt} $\Gamma_Z$ = 2.49430 GeV \\
  $\mu_R=m_Z$ = 91.1535 GeV  &\hspace{20pt} $m_\tau$ = 1.77686 GeV&\hspace{20pt} $m_W$ = 80.3580 GeV \\
   &\hspace{20pt} $m_t$ = 172.690 GeV &\hspace{20pt} $\Gamma_W$ = 2.08400 GeV \\ 
   &\hspace{20pt} &\hspace{20pt} $m_H$ = 125.250 GeV \\
   \hline\hline
 \end{tabular} 
\end{center}
\begin{table}[t]
    \centering
    \begin{tabular}{c|c} 
    \hline
        \multirow{3}{*}{$\epsilon^{-3}$} 
        & $-35.4668 $  \\
        & $\phantom{-}35.4672$ \\
        &0 \\
    \hline
        \multirow{3}{*}{$\epsilon^{-2}$} 
         &  $\phantom{-}1 \ 856.03 + 374.962\ i $ \\
         & $-1\ 207.81 + 241.417 \ i \ $ \\
         & $ \ \ -648.21 - 616.378 \ i $ \\
    \hline
        \multirow{3}{*}{$\epsilon^{-1}$} 
        & $-35 \ 083.52 - \phantom{1}4\ 774.08\ i$ \\
        & $\phantom{-3}9 \ 463.56 - \phantom{1}5\ 495.62\ i$ \\
        & $\phantom{-}25\ 620.07 + 10\ 269.72\ i$\\
    \hline
        \multirow{3}{*}{$\epsilon^{0}$} 
        &$-523\ 472.8 + \phantom{1}82\ 010.2\ i$  \\
        &$\phantom{-1}32\ 550.2 - \phantom{1}16\ 686.4\ i$ \\
        &$\phantom{-1}36\ 267.7 + \phantom{1}52\ 963.0\ i$  \\
    \hline 
        {$\langle {\cal M}^{(0,0)}|{\cal M}^{(0,2),fin}\rangle$}
        & $-454\ 654.8 + 118\ 286.9\ i$ \\
    \hline
    \end{tabular}
    \caption{Numerical coefficient of the IR poles present in the different contributions to Equation~(\ref{eq:subtracted}), in units of $\alpha^4$, for $s=10000 \ \text{GeV}^2$ and $t=-2500 \ \text{GeV}^2$. For each pole, the first row refers to $\langle {\cal M}^{(0,0)}|{\cal M}^{(0,2)}_{\rm fer}\rangle$, the second one to $-{\cal I}^{(0,2)}_{\rm fer}\;\langle {\cal M}^{(0,0)}|{\cal M}^{(0,0)}\rangle$ and the third one to $-{\cal I}^{(0,1)} \; \langle {\cal M}^{(0,0)}|{\cal M}^{(0,1),fin}_{\rm fer}\rangle$. The last row is the final result of $\langle {\cal M}^{(0,0)}|{\cal M}^{(0,2),fin}_{\rm fer}\rangle$.}
    \label{tab:ircanc}
\end{table}

\subsection{Separately UV finite contributions}
\label{sec:UVfinite}
The cancellation of all the poles in $\epsilon$ takes place with the combination of three independent calculations: the diagrammatic part, the UV counterterm diagrams, the IR subtraction. The diagrammatic part features process dependent Feynman integrals and amplitude structures. The UV counterterms and the IR subtraction term are instead process independent and can be used in the evaluation of any scattering process. 
The factorisation of the IR divergences is demonstrated on general grounds and makes the IR subtraction contribution an independent element that we combine with the other two. We stress that there is no additional freedom that can be exploited to obtain the cancellation of the poles, such that the achievement of the latter qualifies as the simultaneous check of UV  and IR finiteness.

A detailed inspection of the cancellations of the UV divergences within specific subsets of contributions can also be performed, partially relying on the BFG quantisation \cite{Denner:1994xt}.
We illustrate a few examples where specific patterns have been observed:
\begin{itemize}
    \item We observe that all the corrections due to the exchange of neutral bosons (photon, $Z$, Higgs, the neutral pseudoscalar) correct the quark vertex function (and separately the muon vertex function) and satisfy a QED-like Ward identity, when combined with the corresponding contributions to the external fermion WF  renormalisation constants. The UV finiteness of such combinations (vertex and WF corrections) is not trivial and holds at arbitrary virtualities.
    \item The one-loop vacuum polarisation of a massive gauge boson has a transverse and a longitudinal parts. We consider e.g. the insertion in a one-loop massive box diagram  of a fermionic self-energy (e.g. Figure-\ref{fig:sample2Ldiagrams}-(k)) and we combine it with the one-loop renormalisation (gauge-boson mass and charge on the adjacent vertices) of the underlying one-loop diagram (e.g. Figure \ref{fig:fermionloopandCTs}-(c)). The one-loop charge renormalisation is a constant which multiplies the whole one-loop diagram and respects its tensor structure: e.g. in 't Hooft-Feynman gauge the gauge boson propagator is proportional to the metric tensor. We have checked that, after the UV sub-renormalisation, the combination of the two-loop diagram and its one-loop sub-renormalisation terms still features a UV divergence due to the longitudinal contributions of the fermionic self-energy insertion. The latter cancels at the level of the complete amplitude, thanks to the overall symmetries of the latter. In the case of two-loop boxes with neutral gauge boson exchanges, we observe the exact cancellation of this UV pole when we sum the planar and the crossed topologies (Figure \ref{fig:sample2Ldiagrams}-(k,l)). In the case of two-loop boxes with $W$ boson exchanges, since only the planar configuration is allowed by electric charge conservation, the longitudinal UV divergences have to be combined with analogous contributions from the vertex (Figure \ref{fig:sample2Ldiagrams}-(e-j)), self-energy (Figure \ref{fig:sample2Ldiagrams}-(a-d)), and WF corrections, eventually obtaining a cancellation. In the case of the vertex corrections with neutral gauge-boson exchanges, this cancellation goes along with the other ones described in the previous point.
    \item The vertex corrections with a fermionic triangle are connected with the external fermion line either with two neutral or with two $W$ gauge bosons. In the former case we observe a cancellation of the UV poles within this subset, whereas in the latter case they contribute to the cancellation of the UV poles which appear in the Feynman diagrams with the exchange of charged $W$ bosons.
\end{itemize}
The cancellation of the UV divergences for the diagram contributions featuring exchange of neutral gauge bosons has been deduced a posteriori after the subtraction of the IR divergences.
The latter can be systematically verified looking for specific subsets of diagrams according to their proportionality with the electric charges $Q_u, Q_\mu, Q_f$.

\subsection{Independence of the \texorpdfstring{$\gamma_5$}{gamma5} prescription}
\label{sec:gammafiveindep}
The loss of the trace cyclicity is addressed in {\sc Abiss} by leaving to the user the possibility of choosing which  point will be considered as starting point for the ordering of the Dirac matrices. We consider two cases:
\begin{itemize}
    \item 
We define fixed-point convention the choice that realises the desired external spinors ordering in the sum over the polarisation states.
While each choice leads to different intermediate results at the level of single interference terms, this dependence always cancels in the SM when combining together all the diagrammatic contributions.
 \item 
A second convention  is the choice of the reading point for the internal fermionic traces, like those of the  topologies (g) and (j) of Figure \ref{fig:sample2Ldiagrams}.
For SM computations, the independence from the reading point is retained independently for each of the three fermion families only after summing over all the respective diagrammatic contributions.

\end{itemize}

As a consistency check of our procedure, we have performed the full computation with all possible choices for the fixed-point and reading point conventions, verifying that our results are independent of any of these prescriptions.

\subsection{Impact of the complex mass scheme on the cross section evaluation}

In the classical Lagrangian, the bare parameters are real valued, in order to fulfill the unitarity constraint.
This feature has an impact in the CMS on the value of the renormalised parameters, which acquire an imaginary part identical in absolute value and of opposite sign with respect to the one of their associated counterterms.
In the case of the masses, the imaginary part has a possible interpretation as the decay width of the particle; in the case of the electric charge on the contrary, the renormalisation condition at $k^2=0$, below any physical threshold, clashes with the presence of an imaginary part. The fact that the renormalised parameters are complex valued is just a sign of the rearrangement of higher-order corrections during the renormalisation step.

We present here a few comments.
\begin{itemize}
    \item 
In the CMS the definition of the renormalised mass of all particles, also the stable ones like the electron, has to be implemented in the complex plane, in order to respect the gauge invariance of the physical amplitude. 

\item
The imaginary part of the complex-valued squared mass is not an independent parameter, but on the contrary it can be computed from first principles in terms of the basic inputs of the Lagrangian, as ${\rm Im}(\mu_V^2)=-{\rm Im}\left(\Sigma_{VV}(\mu_V^2) \right)$.
In order to preserve the \oaa\,accuracy of the calculation also at the resonance of the gauge boson, the imaginary part of the self-energy at \oaaa\, is needed.

Using the parameterisation $\mu_V^2=m_V^2-i\,m_V \Gamma_V$ with $\Gamma_V$ the experimental width breaks the SM prediction by treating the imaginary part of the gauge boson mass as an independent parameter, and it
introduces formal inaccuracies, at the same order of the calculation.
We observe that indeed ${\rm Im}\left(\Sigma_{VV}(\mu_V^2) \right)-{\rm Im}\left(\Sigma_{VV}(m_V^2) \right)=
{\cal O}(\alpha^2)$,
where ${\rm Im}\left(\Sigma_{VV}(m_V^2) \right)$ is related to the standard definition of decay width $\Gamma$.

\item
A complex-valued renormalised electric charge is not an obstacle in the preparation of the finite UV-renormalised and IR-subtracted amplitude, because all the manipulations take place at the amplitude level and the overall coupling constant can be factored out from all the terms.
An inconsistency appears,
as discussed in \cite{Buccioni:2019sur,Frederix:2018nkq},
when we combine the two-loop virtual corrections (proportional to $\alpha^3\alpha^*$) with the double-real (proportional to $\alpha^2(\alpha^*)^2$) and real-virtual ones (proportional to $\alpha^{5/2} (\alpha^*)^{3/2}$).
A prescription for the evaluation of the total cross section will be discussed in a separate publication.
\end{itemize}

\section{Conclusions}
\label{sec:conclusions}
We have presented in this paper the evaluation of the complete set of fermionic two-loop EW virtual corrections to the process of  production of a massive lepton-pair, in quark-antiquark annihilation.

These results illustrate the solution to several major conceptual and technical problems relevant in the study of EW radiative corrections: the renormalisation program, the cancellation of the IR divergences, the independence from the prescription needed to handle $\gamma_5$, the evaluation of the Feynman integrals with several internal massive lines.
The gauge invariance of the results is enforced relying on the CMS for the mass renormalisation of the unstable gauge bosons and by requiring the verification of all the relevant bare and renormalised Ward identities satisfied by the Green's function.
The universality of several ingredients of this calculation opens the way to the evaluation of analogous sets of radiative corrections to other scattering processes.

The reshuffling of higher-order effects implied by the CMS leads to potential violations of perturbative unitarity, which will be addressed in a separate publication devoted to the study of the physical cross section.

\section*{Acknowledgments}
We would like to thank Ayres Freitas for several interesting discussions and for the comparison of numerical results for the fermionic triangular contributions.
T.A. is a Research Fellow of the Fonds de la Recherche Scientifique – FNRS.
The work of S.D. received support from the European Union (ERC, MultiScaleAmp, Grant Agreement No. 101078449) and the FWO (contract No. 1227426N). Views and opinions expressed are however those of the author(s) only and do not necessarily reflect those of the European Union or the European Research Council Executive Agency. Neither the European Union nor the granting authority can be held responsible for them.

\newpage
\appendix

\section{Semi-analytical evaluation of the renormalisation constants at \texorpdfstring{$k^2=0$}{k2=0}}
\label{app:semianev}

The determination of the SM counterterms at \oaa requires the evaluation of both the self-energies and their derivatives with respect to their external momentum $k^2$ in correspondence of the threshold values $k^2=m^2$. While in Section \ref{sec:OS_limits} we described our general procedure to consistently perform the on-shell limits, we focus here on the specific case of $k^2\to0$, needed for the photon and massless fermion self-energies $\Sigma^{\hat{A}\hat{A}}_T$ and $\Sigma^{u\bar{u}}_{(L,R,S)}$:
\begin{itemize}
    \item $\Sigma^{\hat{A}\hat{A} (2)}_T(k^2)$ presents MIs coefficients that start from $(k^2)^{-1}$ and $\epsilon^0$ in the respective expansions. This self-energy has to be derived, so the MIs are needed up to $\mathcal{O}((k^2)^2)$ and $\mathcal{O}(\epsilon^0)$;
    \item $\Sigma^{u\bar{u} (2)}_{(L,R,S)}(k^2)$ presents MIs coefficients that start from $(k^2)^{-2}$ and $\epsilon^{-1}$, but in this case the derivative of the self-energy is not needed because $m_u=0$ (see Equation (\ref{eq:deltaZf2})). Consequently, the MIs are needed up to $\mathcal{O}((k^2)^2)$ and $\mathcal{O}(\epsilon)$.
\end{itemize} 

The limit $k^2\to0$ does not commute with the $\epsilon$ expansion in dimensional regularisation. Taking the on-shell limit as a first step encodes the IR divergences as $\epsilon$ poles in the self-energy expressions, and prevents the appearance of logarithmic terms $\log(k^2)$. Hence, we first expand every master integral $J_i(k^2)$ up to $\mathcal{O}((k^2)^2)$ according to:
\begin{equation}
    J_i(k^2) = J_i(0) + k^2 \dfrac{\partial J_i(k^2)}{\partial k^2}\bigg|_{0} + \dfrac{1}{2!} (k^2)^2 \dfrac{\partial^2 J_i(k^2)}{\partial (k^2)^2}\bigg|_{0} + \mathcal{O}((k^2)^3)
\end{equation}
The problem has now moved to the evaluation of $J_i(0)$ and its derivatives in $k^2=0$. The latter can be reduced to a combination of the master integrals $\{J_1(0), \dots , J_n(0) \}$  of the same  family of $J_i(k^2)$, through a recursive application of differential equations and on-shell reduction rules obtained with standard algorithms \cite{Lee:2013mka,Lange:2025fba}.

\begin{figure}
\centering
\begin{tikzpicture}[thick]
    \begin{scope}[xshift=0cm]
        \draw (-3.5, 0) -- (-1.5, 0);
        \draw (1.5, 0) -- (3.5, 0);

        \draw (-1.5, 0) arc (180:135:1.5);
        \draw (135:1.5) arc (135:45:1.5);
        \draw (45:1.5) arc (45:0:1.5);
        \draw (-1.5, 0) arc (180:360:1.5);

        \draw (135:1.5) arc (225:315:1.5);

        \draw[-stealth, thin] (-3.0, 0.2) -- (-2.2, 0.2);
        \node at (-2.6, 0.45) {\small $k$};

        \draw[-stealth, thin] (2.2, 0.2) -- (3.0, 0.2);
        \node at (2.6, 0.45) {\small $k$};

        \draw[-stealth, thin] (-1.8, 0.3) -- (-1.3, 1.2);
        \node at (-1.75, 0.85) {\small $p$};

        \draw[-stealth, thin] (1.3, 1.2) -- (1.8, 0.3);
        \node at (1.75, 0.85) {\small $p$};

        \draw[-stealth, thin] (-0.4, 1.75) -- (0.4, 1.75);
        \node at (0, 2.0) {\small $q$};

        \draw[-stealth, thin] (-0.4, 0.45) -- (0.4, 0.45);
        \node at (0, 0.2) {\small $p-q$};

        \draw[-stealth, thin] (-1.9, -0.6) -- (-1.4, -1.3);
        \node at (-2.15, -1.0) {\small $k-p$};
    \end{scope}

    \begin{scope}[xshift=6.5cm]
        \draw (0,0) circle (1.5);

        \draw (0, 1.5) -- (0, -1.5);

        \draw[-stealth, thin] (-1.6, 0.6) -- (-0.9, 1.4);
        \node at (-1.5, 1.2) {\small $p$};

        \draw[-stealth, thin] (0.9, 1.4) -- (1.6, 0.6);
        \node at (1.5, 1.2) {\small $q$};

        \draw[-stealth, thin] (-0.2, 0.5) -- (-0.2, -0.5);
        \node at (-0.7, 0) {\small $p-q$};
    \end{scope}

\end{tikzpicture}
\caption{\label{fig:Davydychev_integrals} Self-energy and vacuum diagrams connected through the mathematical limit of external $k^2 \rightarrow 0$ \cite{Davydychev:1992mt}.}
\end{figure}
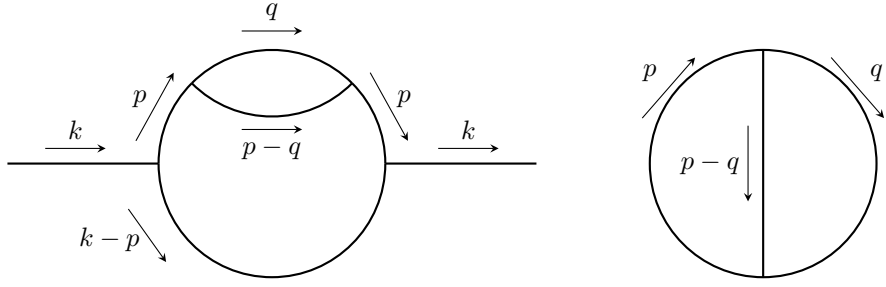

The evaluation of $\{J_1(0), \dots , J_n(0) \}$ keeping the exact $\epsilon$-dependence has been performed following the strategy presented in \cite{Davydychev:1992mt}. Every $J_i(0)$ has been expressed as a combination of vacuum integrals, after applying the decomposition formula 
\begin{equation} \label{eq:Davydychev 2.2}
    \dfrac{1}{(p^2-m_i^2)(p^2-m_j^2)} = \dfrac{1}{m_i^2-m_j^2} \left( \dfrac{1}{p^2-m_i^2} - \dfrac{1}{p^2-m_j^2}\right) \ ,
\end{equation} 
and consistently performing the $k^2\to0$ limit inside the integrand. In some cases, this procedure required additional shifts on the integration variables, in order to ensure the presence of a mass term in all the denominators featuring a dependence on the external momentum $k$. In addition, the two-loop integrals featuring only two propagators always factorised in the product of two one-loop integrals. 
 A graphical example of the limit $k^2\to 0$ is sketched in Figure \ref{fig:Davydychev_integrals}.

The analytical results for vacuum integrals presented  in \cite{Davydychev:1992mt} are generally valid for real-valued internal masses, and for a computation within the CMS they must be analytically continued in the complex plane. Every term featuring a dependence on $\log(z)$, $\log(1 \pm z)$, $\text{Li}_2(z)$ or $\arctan(z)$, with $z$ being a complex-valued ratio involving at least one of $\mu_W$ and $\mu_Z$, has been treated according to the Riemann sheet prescription presented in Section \ref{sec:OS_limits}. This choice preserves the causality of the final expressions, that in Real Mass Scheme is ensured by adopting the Feynman prescription. Only at this stage, the analytical expressions of the MIs in terms of hypergeometric functions have been expanded in series of $\epsilon$ and used for the determination of the \oaa counterterms.

\bibliography{long}
\bibliographystyle{JHEP}

\end{document}